\documentclass[twocolumn,floatfix,showpacs,showkeys,preprintnumbers,pre,amssymb,aps,a4paper]{revtex4} 

\usepackage{amsmath,amsfonts,amssymb}
\usepackage{array}
\usepackage{bm}
\usepackage{cancel}
\usepackage{dcolumn}
\usepackage{graphicx}
\usepackage{picins}
\usepackage[scanall]{psfrag}
\usepackage{times}

\usepackage[
	paperwidth=210mm,
	paperheight=297mm,
	textwidth=170mm,
	textheight=250mm,
	dvips]
{geometry}

\newcommand{\psfragstyle}[1]{\small{#1}}

\newcommand{\figref}[1]{Fig.~\ref{#1}}

\newcommand{\tableref}[1]{Table~\ref{#1}}

\newlength{\tempfboxsep}
\newcommand{\sidebar}[1]{%
	\setlength{\fboxsep}{0pt}%
	\setlength{\temptabcolsep}{\tabcolsep}%
	\setlength{\tabcolsep}{0pt}%
	\begin{tabular}{rcl}%
		\vrule width 0.03\columnwidth &%
		\hspace*{0.05\columnwidth} &%
		\begin{minipage}{0.92\columnwidth}%
			#1%
		\end{minipage}%
	\end{tabular}%
	\setlength{\tabcolsep}{\temptabcolsep}%
	\setlength{\fboxsep}{\tempfboxsep}%
}
\newlength{\temptabcolsep}

\newcommand{\zero}{\ensuremath{\text{0}}}
\newcommand{\one}{\ensuremath{\text{1}}}
\newcommand{\two}{\ensuremath{\text{2}}}
\newcommand{\three}{\ensuremath{\text{3}}}
\newcommand{\four}{\ensuremath{\text{4}}}

\newcommand{\ith}{\ensuremath{i^{\text{th}}}~}

\newcommand{\theth}{\ensuremath{^{\text{th}}}~}

\newcommand{\trademark}{\ensuremath{^{\text{\tiny{TM}}}}~}

\begin{document}

\newlength{\figurewidth}
\setlength{\figurewidth}{\columnwidth}
\newlength{\halffigurewidth}
\setlength{\halffigurewidth}{0.47\columnwidth}
\newlength{\figureseparation}
\setlength{\figureseparation}{0.2cm}

\setlength{\parindent}{0pt}
\setlength{\parskip}{1ex plus 0.5ex minus 0.2ex}

\title{
  Traffic Flow Theory
}

\author{Sven Maerivoet}
  \email{sven.maerivoet@esat.kuleuven.be}
\author{Bart De Moor}
\affiliation{
  Department of Electrical Engineering ESAT-SCD (SISTA)\footnote{
    Phone: +32 (0) 16 32 17 09 Fax: +32 (0) 16 32 19 70\\
    URL: \texttt{http://www.esat.kuleuven.be/scd}
  }, Katholieke Universiteit Leuven\\
  Kasteelpark Arenberg 10, 3001 Leuven, Belgium
}

\date{\today}

\begin{abstract}
	The scientific field of traffic engineering encompasses a rich set of 
	mathematical techniques, as well as researchers with entirely different 
	backgrounds. This paper provides an overview of what is currently the 
	state-of-the-art with respect to traffic flow theory. Starting with a brief 
	history, we introduce the microscopic and macroscopic characteristics of 
	vehicular traffic flows. Moving on, we review some performance indicators that 
	allow us to assess the quality of traffic operations. A final part of this 
	paper discusses some of the relations between traffic flow characteristics, 
	i.e., the fundamental diagrams, and sheds some light on the different points 
	of view adopted by the traffic engineering community.
\end{abstract}



\preprint{SISTA Internal Report 05-}

\maketitle

\setlength{\parskip}{0pt}

\tableofcontents

\setlength{\parskip}{1ex plus 0.5ex minus 0.2ex}

%

The scientific field of traffic engineering encompasses a rich set of 
mathematical techniques, as well as researchers with entirely different 
backgrounds. This paper provides an overview of what is currently the 
state-of-the-art with respect to traffic flow theory. Starting with a brief 
history, we introduce the microscopic and macroscopic characteristics of 
vehicular traffic flows. Moving on, we review some performance indicators that 
allow us to assess the quality of traffic operations. A final part of this paper 
discusses some of the relations between traffic flow characteristics, i.e., the 
fundamental diagrams, and sheds some light on the different points of view 
adopted by the traffic engineering community.

Because of the large diversity of the scientific field (engineers, physicists, 
mathematicians, \ldots all lack a unified standard or convention), one of the 
principal aims of this paper is to define both a \emph{logical and consistent 
terminology and notation}. It is our strong belief that such a consistent 
notation is a necessity when it comes to creating order in the `zoo of 
notations' that in our opinion currently exists. For a concise but complete 
overview of all abbreviations and notations proposed and adopted throughout this 
paper, we refer the reader to appendix \ref{appendix:Glossary}.

	\section{A brief history of traffic flow theory}

Historically, traffic engineering got its roots as a rather practical 
discipline, entailing most of the time a common sense of its practitioners to 
solve particular traffic problems. However, all this changed at the dawn of the 
1950s, when the scientific field began to mature, attracting engineers from all 
sorts of trades. Most notably, John Glen Wardrop instigated the evolving 
discipline now known as traffic flow theory, by describing traffic flows using 
mathematical and statistical ideas \cite{WARDROP:52}.

During this highly active period, mathematics established itself as a solid 
basis for theoretical analyses, a phenomenon that was entirely new to the 
previous, more `rule-of-thumb' oriented, line of reasoning. Two examples of the 
progress during this decade, include the fluid-dynamic model of Michael James 
Lighthill, Gerald Beresford Whitham, and Paul Richards (or the \emph{LWR model} 
for short) for describing traffic flows \cite{LIGHTHILL:55,RICHARDS:56}, and the 
car-following experiments and theories of the club of people working at General 
Motors' research laboratory \cite{CHANDLER:58,GAZIS:59,HERMAN:59,GAZIS:61}. 
Simultaneous progress was also made on the front of economic theory applied to 
transportation, most notably by the publication of the `BMW trio', Martin J.
Beckmann, Charles Bartlett McGuire, and Christopher B. Winsten 
\cite{BECKMANN:55}.

From the 1960s on, the field evolved even further with the advent of the early 
personal computers (although at that time, they could only be considered as mere 
computing units). More control-oriented methods were pursued by engineers, as a 
means for alleviating congestion at tunnels and intersections, by e.g., 
adaptively steering traffic signal timings. Nowadays, the field has been kindly 
embraced by the industry, resulting in what is called \emph{intelligent 
transportation systems} (ITS), covering nearly all aspects of the transportation 
community.

In spite of the intense booming during the 1950s and 1960s, all progress 
seemingly came to sudden stop, as there were almost no significant results for 
the next two decades (although there are some exceptions, such as the 
significant work of Ilya Prigogine and Robert Herman's, who developed a traffic 
flow model based on a gas-kinetic analogy \cite{PRIGOGINE:71}). One of the main 
reasons for this, stems from the fact that many of the involved key players 
returned to their original scientific disciplines, after exhausting the 
application of their techniques to the transportation problem \cite{NEWELL:02}. 
Note that despite this calm period, the application of control theory to 
transportation started finding new ways to alleviate local congestion problems.

At the beginning of the 1990s, researchers found a revived interest in the field 
of traffic flow modelling. On the one hand, researchers' interests got kindled 
again by the appealing simplicity of the LWR model, whereas on the other hand 
one of the main boosts came from the world of statistical physics. In this 
latter framework, physicists tried to model many particle systems using simple 
and elegant behavioural rules. As an example, the now famous particle hopping 
(cellular automata) model of Kai Nagel and Michael Schreckenberg \cite{NAGEL:92} 
still forms a widely-cited basis for current research papers on the subject.

In parallel with this kind of modelling approach, many of the old `beliefs' 
(e.g., the fluid-dynamic approach to traffic flow modelling) started to get 
questioned. As a consequence, a plethora of models quickly found its way to the 
transportation community, whereby most of these models didn't give a thought as 
to whether or not their associated phenomena corresponded to real-life traffic 
observations.\\

\sidebar{
	We note here that, whatever the modelling approach may be, researchers should 
	always compare their results to the reality of the physical world. Ignoring 
	this basic step, reduces the research in our opinion to nothing more than a 
	mathematical exercise~!
}\\

As the international research community began to spawn its traffic flow 
theories, Robert Herman aspired to bring them all together in december 1959. 
This led to the tri-annual organisation of the \emph{International Symposium on 
Transportation and Traffic Theory} (ISTTT), by some heralded as `the Olympics of 
traffic theory' because the symposium talks about the fundamentals underlying 
transportation and traffic phenomena. Another example of the evolution of recent 
developments with respect to the parallels between traffic flows and granular 
media, is the bi-annual organisation of the workshop on \emph{Traffic and 
Granular Flow} (TGF), a platform for exchanging ideas by bringing together 
researchers from various scientific fields.

Nowadays, the research and application of traffic flow theory and intelligent 
transportation systems continues. The scientific field has been largely 
diversified, encompassing a broad range of aspects related to sociology, 
psychology, the environment, the economy, \ldots The global avidity of the field 
can be witnessed by the exponentially growing publication output. Keeping our 
previous comment in mind, researchers from time to time just seem to `add to the 
noise' (mainly due to the sheer diversity of the literature body), although 
there occasionally exist exceptions such as the late Newell, as subtly pointed 
out by Michael Cassidy in \cite{ORRICK:02}.

As a final word, we refer the reader to two personalised views on the history of 
traffic flow theory, namely the musings of the late Gordon Newell and Denos 
Gazis \cite{NEWELL:02,GAZIS:02}. We furthermore invite the reader to cast a 
glance at the ending pages of Wardrop's paper \cite{WARDROP:52}, in which a 
rather colourful discussion on the introduction of mathematics to traffic flow 
theory has been written down.

	\section{Microscopic traffic flow characteristics}
	\label{sec:TFT:MicroCharacteristics}

Road traffic flows are composed of drivers associated with individual vehicles, 
each of them having their own characteristics. These characteristics are called 
\emph{microscopic} when a traffic flow is considered as being composed of such a 
stream of vehicles. The dynamical aspects of these traffic flows are formed by 
the underlying interactions between the drivers of the vehicles. This is largely 
determined by the behaviour of each driver, as well as the physical 
characteristics of the vehicles.

Because the process of participating in a traffic flow is heavily based on the 
behavioural aspects associated with human drivers \cite{GARTNER:97}, it would 
seem important to include these human factors into the modelling equations. 
However, this leads to a severe increase in complexity, which is not always a 
desired artifact \cite{MAERIVOET:01}. However, in the remainder of this section, 
we always consider a vehicle-driver combination as a single entity, taking only 
into account some vehicle related traffic flow characteristics.\\

\sidebar{
	Note that despite our previous remarks, we do not debate the necessity of a 
	psychological treatment of traffic flow theory. As the research into driver 
	behaviour is gaining momentum, a lot of attention is gained by promising 
	studies aimed towards driver and pedestrian safety, average reaction times, 
	the influence of stress levels, aural and visual perceptions, ageing, medical 
	conditions, fatigue, \ldots
}\\

		\subsection{Vehicle related variables}

Considering individual vehicles, we can say that each vehicle $i$ in a lane of a 
traffic stream has the following informational variables:

\begin{itemize}
	\item a \emph{length}, denoted by $l_{i}$,
	\item a \emph{longitudinal position}, denoted by $x_{i}$,
	\item a \emph{speed}, denoted by $v_{i} = \displaystyle \frac{dx_{i}}{dt}$,
	\item and an \emph{acceleration}, denoted by $a_{i} = \displaystyle \frac{dv_{i}}{dt} =
		\displaystyle \frac{d^{\two}x_{i}}{dt^{\two}}$.
\end{itemize}

Note that the position $x_{i}$ of a vehicle is typically taken to be the 
position of its rear bumper. In this first approach, a vehicle's other spatial 
characteristics (i.e., its width, height, and lane number) are neglected. And in 
spite of our narrow focus on the vehicle itself, the above list of variables is 
also complemented with a driver's \emph{reaction time}, denoted by $\tau_{i}$.

With respect to the acceleration characteristics, it should be noted that these 
are in fact not only dependent on the vehicle's engine, but also on e.g., the 
road's inclination, being a non-negligible factor that plays an important role 
in the forming of congestion at bridges and tunnels. We do not use the 
derivative of the acceleration, called \emph{jerk}, \emph{jolt}, or \emph{surge} 
(jerk is also used to represent the smoothness of the \emph{acceleration noise} 
\cite{MONTROLL:64}).

Except in the acceleration capabilities of a vehicle, we ignore the physical 
forces that act on a vehicle, e.g., the earth's gravitational pull, road and 
wind friction, centrifugal forces, \ldots A more elaborate explanation of these 
forces can be found in \cite{DAGANZO:97}.

		\subsection{Traffic flow characteristics}
		\label{sec:TFT:MicroCharacteristicsForTrafficFlow}

\begin{figure}[!htb]
	\centering
	\psfrag{(i)}[][]{\psfragstyle{$(i)$}}
	\psfrag{(i+1)}[][]{\psfragstyle{$(i + \one)$}}
	\psfrag{xi}[][]{\psfragstyle{$x_{i}$}}
	\psfrag{xi+1}[][]{\psfragstyle{$x_{i + \one}$}}
	\psfrag{li}[][]{\psfragstyle{$l_{i}$}}
	\psfrag{gsi}[][]{\psfragstyle{$g_{s_{i}}$}}
	\psfrag{hsi}[][]{\psfragstyle{$h_{s_{i}}$}}
	\includegraphics[width=\figurewidth]{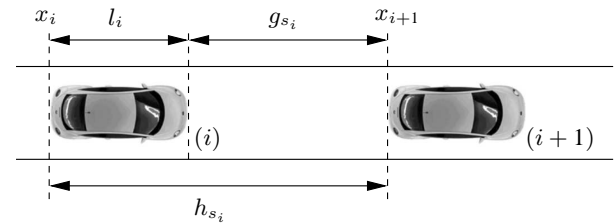}
	\caption{
		Two consecutive vehicles (a follower $i$ at position $x_{i}$ and a leader $i 
		+ \one$ at position $x_{i + \one}$) in the same lane in a traffic stream. 
		The follower has a certain space headway $h_{s_{i}}$ to its leader, equal to 
		the sum of the vehicle's space gap $g_{s_{i}}$ and its length $l_{i}$.
	}
	\label{fig:TFT:SpaceHeadway}
\end{figure}

Referring to \figref{fig:TFT:SpaceHeadway}, we can consider two consecutive 
vehicles in the same lane in a traffic stream: a follower $i$ and its leader $i 
+ \one$. From the figure, it can be seen that vehicle $i$ has a certain 
\emph{space headway} $h_{s_{i}}$ to its predecessor (it is expressed in metres), 
composed of the distance (called the \emph{space gap}) $g_{s_{i}}$ to this 
leader  and its own \emph{length} $l_{i}$:

\begin{equation}
\label{eq:TFT:SpaceHeadway}
	h_{s_{i}} = g_{s_{i}} + l_{i}.
\end{equation}

By taking, as stated before, the rear bumper as a vehicle's position, the space 
headway $h_{s_{i}} = x_{i + \one} - x_{i}$. The space gap is thus measured from 
a vehicle's front bumper to its leader's rear bumper.

Analogously to equation \eqref{eq:TFT:SpaceHeadway}, each vehicle also has a 
\emph{time headway} $h_{t_{i}}$ (expressed in seconds), consisting of a 
\emph{time gap} $g_{t_{i}}$ and an \emph{occupancy time} $\rho_{i}$:

\begin{equation}
\label{eq:TFT:TimeHeadway}
	h_{t_{i}} = g_{t_{i}} + \rho_{i}.
\end{equation}

\begin{figure}[!htb]
	\centering
	\psfrag{time}[][]{\psfragstyle{time}}
	\psfrag{space}[][]{\psfragstyle{space}}
	\psfrag{(i)}[][]{\psfragstyle{$(i)$}}
	\psfrag{(i+1)}[][]{\psfragstyle{$(i + \one)$}}
	\psfrag{xi}[][]{\psfragstyle{$x_{i}$}}
	\psfrag{xi+1}[][]{\psfragstyle{$x_{i + \one}$}}
	\psfrag{ti}[][]{\psfragstyle{$t_{i}$}}
	\psfrag{ti+1}[][]{\psfragstyle{$t_{i + \one}$}}
	\psfrag{hsi}[][]{\psfragstyle{$h_{s_{i}}$}}
	\psfrag{gsi}[][]{\psfragstyle{$g_{s_{i}}$}}
	\psfrag{li}[][]{\psfragstyle{$l_{i}$}}
	\psfrag{hti}[][]{\psfragstyle{$h_{t_{i}}$}}
	\psfrag{gti}[][]{\psfragstyle{$g_{t_{i}}$}}
	\psfrag{rhoi}[][]{\psfragstyle{$\rho_{i}$}}
	\includegraphics[width=\figurewidth]{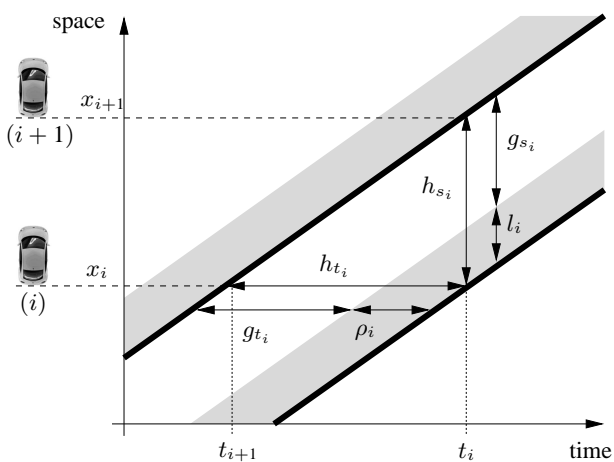}
	\caption{
		A time-space diagram showing two vehicle trajectories $i$ and $i + \one$, as 
		well as the space and time headway $h_{s_{i}}$ and $h_{t_{i}}$ of vehicle 
		$i$. Both headways are composed of the space gap $g_{s_{i}}$ and the vehicle 
		length $l_{i}$, and the time gap $g_{t_{i}}$ and the occupancy time 
		$\rho_{i}$, respectively. The time headway can be seen as the difference in 
		time instants between the passing of both vehicles, respectively at $t_{i + 
		\one}$ and $t_{i}$ (diagram based on \cite{LOGGHE:03}).
	}
	\label{fig:TFT:SpaceAndTimeHeadways}
\end{figure}

Both space and time headways can be visualised in a \emph{time-space diagram}, 
such as the one in \figref{fig:TFT:SpaceAndTimeHeadways}. Here, we have shown 
the two vehicles $i$ and $i + \one$ as they are driving. Their positions $x_{i}$ 
and $x_{i + \one}$ can be plotted with respect to time, tracing out two 
\emph{vehicle trajectories}. As the time direction is horizontal and the space 
direction is vertical, the vehicles' respective speeds can be derived by taking 
the tangents of the trajectories (for simplicity, we have assumed that both 
vehicles travel at the same constant speed, resulting in parallel linear 
trajectories). Accelerating vehicles have steep inclining trajectories, whereas 
those of stopped vehicles are horizontal.

When the vehicle's speed is constant, the time gap is the amount of time 
necessary to reach the current position of the leader when travelling at the 
current speed (i.e., it is the elapsed time an observer at a fixed location 
would measure between the passing of two consecutive vehicles). Similarly, the 
occupancy time can be interpreted as the time needed to traverse a distance 
equal to the vehicle's own length at the current speed, i.e., $\rho_{i} = l_{i} 
/ v_{i}$; this corresponds to the time the vehicle needs to pass the observer's 
location. Both equations \eqref{eq:TFT:SpaceHeadway} and 
\eqref{eq:TFT:TimeHeadway} are furthermore linked to the vehicle's speed $v_{i}$ 
as follows \cite{DAGANZO:97}:

\begin{equation}
\label{eq:TFM:SpaceTimeHeadwayGapSpeedRelation}
	\frac{h_{s_{i}}}{h_{t_{i}}} = \frac{g_{s_{i}}}{g_{t_{i}}} = \frac{l_{i}}{\rho_{i}} = v_{i}.
\end{equation}

As the above definitions deal with what is called single-lane traffic, we can 
easily extend them to multi-lane traffic. In this case, four extra space gaps 
--- related to the vehicles in the neighbouring lanes --- are introduced, namely 
$g_{s_{i}}^{l,f}$ at the left-front, $g_{s_{i}}^{l,b}$ at the left-back, 
$g_{s_{i}}^{r,f}$ at the right-front, and $g_{s_{i}}^{r,b}$ at the right-back. 
The four corresponding space headways, $h_{s_{i}}^{l,f}$, $h_{s_{i}}^{l,b}$, 
$h_{s_{i}}^{r,f}$, and $h_{s_{i}}^{r,b}$, are introduced in a similar fashion. 
The extra time gaps and headways are derived in complete analogy, leading to the 
four time gaps $g_{t_{i}}^{l,f}$, $g_{t_{i}}^{l,b}$, $g_{t_{i}}^{r,f}$, and 
$g_{t_{i}}^{r,b}$, and the four corresponding time headways $h_{t_{i}}^{l,f}$, 
$h_{t_{i}}^{l,b}$, $h_{t_{i}}^{r,f}$, and $h_{t_{i}}^{r,b}$.

In single-lane traffic, vehicles always keep their relative order, a principle 
sometimes called \emph{first-in, first-out} (FIFO) \cite{DAGANZO:95c}. For 
multi-lane traffic however, this principle is no longer obeyed due to overtaking 
manoeuvres, resulting in vehicle trajectories that cross each other. If the same 
time-space diagram were to be drawn for only one lane (in multi-lane traffic), 
then some vehicles' trajectories would suddenly appear or vanish at the point 
where a lane change occurred.\\

\sidebar{
	In some traffic flow literature, other nomenclature is used: \emph{space} for 
	the space headway, \emph{distance} or \emph{clearance} for the space gap, and 
	\emph{headway} for the time headway. Because this terminology is confusing, we 
	propose to use the unambiguously defined terms as described in this section.
}\\

	\section{Macroscopic traffic flow characteristics}
	\label{sec:TFT:MacroCharacteristics}

\begin{figure}[!htb]
	\centering
	\psfrag{t}[][]{\psfragstyle{$t$}}
	\psfrag{x}[][]{\psfragstyle{$x$}}
	\psfrag{dt}[][]{\psfragstyle{$dt$}}
	\psfrag{dx}[][]{\psfragstyle{$dx$}}
	\psfrag{Rt}[][]{\psfragstyle{$R_{t}$}}
	\psfrag{Rs}[][]{\psfragstyle{$R_{s}$}}
	\psfrag{Rts}[][]{\psfragstyle{$R_{t,s}$}}
	\psfrag{Tmp}[][]{\psfragstyle{$T_{\text{mp}}$}}
	\psfrag{K}[][]{\psfragstyle{$K$}}
	\includegraphics[width=\figurewidth]{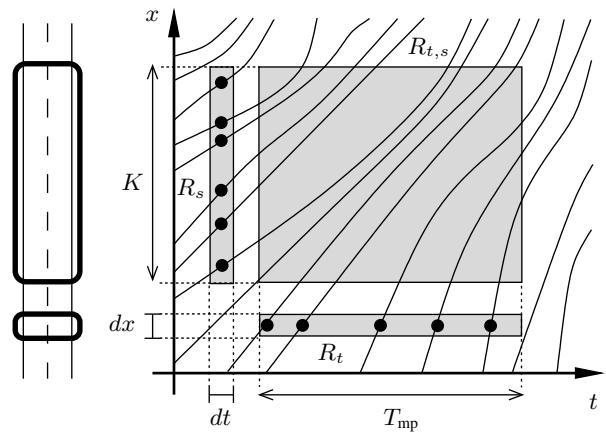}
	\caption{
		A time-space diagram showing several vehicle trajectories and three 
		measurement regions $R_{t}$, $R_{s}$, and $R_{t,s}$. These rectangular 
		regions are bounded in time and space by a measurement period 
		$T_{\text{mp}}$ and a road section of length $K$. The black dots represent 
		the individual measurements.
	}
	\label{fig:TFT:TXDiagramMeasurements}
\end{figure}

When considering many vehicles simultaneously, the time-space diagram mentioned 
in section \ref{sec:TFT:MicroCharacteristicsForTrafficFlow} can be used to 
faithfully represent all traffic. In \figref{fig:TFT:TXDiagramMeasurements} we 
show the evolution of the system, as we have traced the trajectories of all the 
individual vehicles' movements. This time-space diagram therefore provides a 
complete picture of all traffic operations that are taking place (accelerations, 
decelerations, \ldots).

Instead of considering each vehicle in a traffic stream individually, we now 
`zoom out' to a more aggregate \emph{macroscopic} level (traffic streams are 
regarded e.g., as a fluid). In the remainder of this section, we will measure 
some macroscopic traffic flow characteristics based on the shown time-space 
diagram. To this end, we define three measurement regions:

\begin{itemize}
	\item $R_{t}$ corresponding to measurements at a single fixed location in 
	space ($dx$), during a certain time period $T_{\text{mp}}$. An example of this 
	is a single inductive loop detector (SLD) embedded in the road's concrete.
	\item $R_{s}$ corresponding to measurements at a single instant in time 
	($dt$), over a certain road section of length $K$. An example of this is an 
	aerial photograph.
	\item $R_{t,s}$ corresponding to a general measurement region. Although it can 
	have any shape, in this case we restrict ourselves to a rectangular region in 
	time and space. An example of this is a sequence of images made by a video 
	camera detector.
\end{itemize}

With respect to the size of these measurement regions, some caution is advised: 
a too large measurement region can mask certain effects of traffic flows, 
possibly ignoring some of the dynamic properties, whereas a too small 
measurement region may obstruct a continuous treatment, as the discrete, 
microscopic nature of traffic flows becomes apparent.

Using these different methods of observation, we now discuss the measurement of 
four important macroscopic traffic flow characteristics: density, flow, 
occupancy, and mean speed. We furthermore give a short discussion on the moving 
observer method and the use of floating car data.\\

\sidebar{
	With respect to some naming conventions on roadways, two different `standards' 
	exist for some of the encountered terminology, namely the American and the 
	British standard. Examples are: the classic multi-lane high-speed road with	
	on- and off-ramps, which is called a \emph{freeway} or a \emph{super highway} 
	(American), or an \emph{arterial} or \emph{motorway} (British). A main road 
	with intersections is called an \emph{urban highway} (American) or a 
	\emph{carriageway} (British). In this dissertation, we have chosen to adopt 
	the British standard. Finally, in contrast to Great Britain and Australia, we 
	assume that for low-density traffic, everybody drives on the right instead of 
	the left lane.
}\\

		\subsection{Density}
		\label{sec:TFT:Density}

The macroscopic characteristic called \emph{density} allows us to get an idea of 
how crowded a certain section of a road is. It is typically expressed in number 
of vehicles per kilometre (or mile). Note that the concept of density totally 
ignores the effects of traffic composition and vehicle lengths, as it only 
considers the abstract quantity `number of vehicles'.

Because density can only be \emph{measured} in a certain spatial region (e.g., 
$R_{s}$ in \figref{fig:TFT:TXDiagramMeasurements}), it is \emph{computed} for 
temporal regions such as region $R_{t}$ in 
\figref{fig:TFT:TXDiagramMeasurements}. When density can not be exactly measured 
or computed, or when density measurements are faulty, it has to be 
\emph{estimated}. To this end, several available techniques exist e.g., based on 
explicit simulation using a traffic flow propagation model \cite{MUNOZ:03}, 
based on a vehicle reidentification system \cite{COIFMAN:03b}, based on a 
complete traffic state estimator using an extended Kalman filter \cite{WANG:03}, 
or based on a non-linear adaptive observer \cite{ALVAREZ:04}, \ldots

		\subsubsection{Mathematical formulation}
		\label{sec:TFT:DensityMathematicalFormulation}

Using the spatial region $R_{s}$, the density $k$ for single-lane traffic is 
defined as:

\begin{equation}
\label{eq:TFT:DensityRegionSSingleLane}
	k = \frac{N}{K},
\end{equation}

with $N$ the number of vehicles present on the road segment. If we consider 
multi-lane traffic, we have to sum the partial densities $k_{l}$ of each of the 
$L$ lanes as follows:

\begin{equation}
\label{eq:TFT:DensityRegionSMultiLane}
	k = \sum_{l = \one}^{L} k_{l} = \frac{\one}{K} \sum_{l = \one}^{L} N_{l},
\end{equation}

in which $N_{l}$ now denotes the number of vehicles present in lane $l$ 
(equation \eqref{eq:TFT:DensityRegionSMultiLane} is \emph{not} the same as 
averaging over the partial densities of each lane)\footnote{Note that when 
calculating the total density using equation 
\eqref{eq:TFT:DensityRegionSMultiLane}, the partial densities can also 
correspond without loss of generality to different vehicle classes instead of 
just different lanes \cite{WARDROP:52,DAGANZO:97}.}.

In general, density can be defined as \emph{the total time spent by all the 
vehicles in the measurement region, divided by the area of this region} 
\cite{EDIE:65,DAGANZO:97}. This generalisation allows us to compute the density 
at a point using the measurement region $R_{t}$:

\begin{equation}
\label{eq:TFT:DensityRegionTSingleLane}
	k = \frac{\displaystyle \sum_{i = \one}^{N} T_{i}}{T_{\text{mp}}~dx}
	  = \frac{\one}{T_{\text{mp}}~\cancel{dx}} \sum_{i = \one}^{N} \frac{\cancel{dx}}{v_{i}}
	  = \frac{\one}{T_{\text{mp}}} \sum_{i = \one}^{N} \frac{\one}{v_{i}},
\end{equation}

with $T_{i}$ the travel time and $v_{i}$ the speed of the \ith vehicle. 
Extending the previous derivation to multi-lane traffic is done straightforward 
using equation \eqref{eq:TFT:DensityRegionSMultiLane}:

\begin{equation}
\label{eq:TFT:DensityRegionTMultiLane}
	k = \frac{\one}{T_{\text{mp}}} \sum_{l = \one}^{L} \sum_{i = \one}^{N_{l}} \frac{\one}{v_{i,l}},
\end{equation}

with now $v_{i,l}$ denoting the speed of the \ith vehicle in lane $l$.

As we now can obtain the density in both spatial and temporal regions, $R_{s}$ 
and $R_{t}$ respectively, it would seem a logical extension to find the density 
in the region $R_{t,s}$. In order to do this, however, we need to know the 
travel times $T_{i}$ of the individual vehicles, as can be seen in equation 
\eqref{eq:TFT:DensityRegionTSingleLane}. Because this information is not always 
available, and in most cases rather difficult to measure, we use a different 
approach, corresponding to the temporal average of the density. Assuming that at 
each time step $t$, during a certain time period $T_{\text{mp}}$, the density 
$k(t)$ is known in consecutive regions $R_{s}$, the generalised definition leads 
to the following formulation:

\begin{equation}
\label{eq:TFT:DensityAverageFormulation}
	k =
		\left \lbrace
			\begin{array}{ll}
				\displaystyle \frac{\one}{T_\text{mp}} \int_{t = \zero}^{T_\text{mp}} k(t)~dt & \text{(continuous)},\\
				& \\
				\displaystyle \frac{\one}{T_\text{mp}} \sum_{t = \one}^{T_\text{mp}} k(t) & \text{(discrete)}.
			\end{array}
		\right.
\end{equation}

For multi-lane traffic, combining equations 
\eqref{eq:TFT:DensityRegionSMultiLane} and 
\eqref{eq:TFT:DensityAverageFormulation} results in the following formula for 
computing the density in region $R_{t,s}$ using measurements in discrete time:

\begin{equation}
	k = \frac{\one}{T_\text{mp}~K} \sum_{t = \one}^{T_\text{mp}} \sum_{l = \one}^{L} N_{l}(t),
\end{equation}

where $N_{l}(t)$ denotes the number of vehicles present in lane $l$ at time $t$.

There exists a relation between the macroscopic traffic flow characteristics and 
those microscopic characteristics defined in section 
\ref{sec:TFT:MicroCharacteristicsForTrafficFlow}. For the density $k$, this 
relation is based on the average space headway $\overline h_{s}$ 
\cite{WARDROP:52,DAGANZO:97}:

\begin{equation}
\label{eq:TFT:DensitySpaceHeadwayRelation}
	k = \frac{N}{K}
	  = \frac{N}{\displaystyle \sum_{i = \one}^{N} h_{s_{i}}}
	  = \frac{\one}{\frac{\one}{N} \displaystyle \sum_{i = \one}^{N} h_{s_{i}}}
	  = \frac{\one}{\overline h_{s}},
\end{equation}

with ${\overline h_{s}}^{-\one}$ the reciprocal of the average space headway.

			\subsubsection{Passenger car units}
			\label{sec:TFT:PCUs}

When considering heterogeneous traffic flows (i.e., traffic streams composed of 
different types of vehicles), operating agencies usually don't express the 
macroscopic traffic flow characteristics using the raw number of vehicles, but 
rather employ the notion of \emph{passenger car units} (PCU). These PCUs, 
sometimes also called \emph{passenger car equivalents} (PCE), try to take into 
account the spatial differences between vehicle types. For example, by denoting 
one average passenger car as 1 PCU, a truck in the same traffic stream can be 
considered as 2 PCUs (or even higher and fractional values for trailer 
trucks).\\

\sidebar{
	Let us finally note that, because density is essentially defined as a spatial 
	measurement, it is one of the most difficult quantities to obtain. It is 
	interesting to notice that at this moment, it is theoretically possible for 
	video cameras to measure density over a short spatial region. However, to our 
	knowledge there currently exists no commercial implementation.
}\\

		\subsection{Flow}
		\label{sec:TFT:Flow}

Whereas density typically is a spatial measurement, \emph{flow} can be 
considered as a temporal measurement (i.e., region $R_{t}$). Flow, which we use 
as a shorthand for rate of flow, is typically expressed as an hourly rate, i.e., 
in number of vehicles per hour. Note that sometimes other synonyms such as 
\emph{intensity}, \emph{flux}, \emph{throughput}, \emph{current}, or 
\emph{volume}\footnote{In most cases, volume denotes the number of vehicles 
counted during a certain time period, as opposed to flow which is just the 
equivalent hourly rate.} are used, typically depending on a person's scientific 
background (e.g., engineering, physics, \ldots).

		\subsubsection{Mathematical formulation}

Measuring the flow $q$ in region $R_{t}$ for single-lane traffic, is done using 
the following equation, which is based on raw vehicle counts:

\begin{equation}
\label{eq:TFT:FlowRegionTSingleLane}
	q = \frac{N}{T_{\text{mp}}},
\end{equation}

with $N$ the number of vehicles that has passed the detector's site. For 
multi-lane traffic, we sum the partial flows of each of the $L$ lanes:

\begin{equation}
\label{eq:TFT:FlowRegionTMultiLane}
	q = \sum_{l = \one}^{L} q_{l} = \frac{\one}{T_{\text{mp}}} \sum_{l = \one}^{L} N_{l},
\end{equation}

with now $N_{l}$ denoting the number of vehicles that passed the detector's site 
in lane $l$. Note that we assume that each lane has its own detector, otherwise 
we would be dealing with an average flow across all the lanes.

Generally speaking, flow can defined as \emph{the total distance travelled by 
all the vehicles in the measurement region, divided by the area of this region} 
\cite{EDIE:65,DAGANZO:97}. In analogy with equation 
\eqref{eq:TFT:DensityRegionTSingleLane}, this generalisation allows us to 
compute the flow using the spatial measurement region $R_{s}$:

\begin{equation}
\label{eq:TFT:FlowRegionSSingleLane}
	q = \frac{\displaystyle \sum_{i = \one}^{N} X_{i}}{K~dt}
	  = \frac{\one}{K~\cancel{dt}} \sum_{i = \one}^{N} v_{i}~\cancel{dt}
	  = \frac{\one}{K} \sum_{i = \one}^{N} v_{i},
\end{equation}

with now $X_{i}$ the distance travelled by the \ith vehicle during the 
infinitesimal time interval $dt$. The extension to multi-lane traffic is 
straightforward:

\begin{equation}
\label{eq:TFT:FlowRegionSMultiLane}
	q = \frac{\one}{K} \sum_{l = \one}^{L} \sum_{i = \one}^{N_{l}} v_{i,l}.
\end{equation}

Considering consecutive flow measurements in region $R_{t,s}$, we can derive a 
formulation corresponding to the temporal average of the flow, similar to that 
of equation \eqref{eq:TFT:DensityAverageFormulation}. Assuming that at each time 
step $t$, during a certain time period $T_{\text{mp}}$, the flow $q(t)$ is known 
in consecutive regions $R_{s}$, the generalised definition leads to the 
following equations:

\begin{equation}
\label{eq:TFT:FlowAverageFormulation}
	q =
		\left \lbrace
			\begin{array}{ll}
				\displaystyle \frac{\one}{T_\text{mp}} \int_{t = \zero}^{T_\text{mp}} q(t)~dt & \text{(continuous)},\\
				& \\
				\displaystyle \frac{\one}{T_\text{mp}} \sum_{t = \one}^{T_\text{mp}} q(t) & \text{(discrete)},
			\end{array}
		\right.
\end{equation}

For multi-lane traffic, combining equations \eqref{eq:TFT:FlowRegionSMultiLane} 
and \eqref{eq:TFT:FlowAverageFormulation} results in the following formula for 
computing the flow in region $R_{t,s}$ using measurements in discrete time:

\begin{equation}
	q = \frac{\one}{T_\text{mp}~K} \sum_{t = \one}^{T_\text{mp}} \sum_{l = \one}^{L} \sum_{i = \one}^{N_{l}(t)} v_{i,l}(t),
\end{equation}

where $v_{i,l}(t)$ denotes the speed of the \ith vehicle in lane $l$ at time 
$t$.

In analogy with equation \eqref{eq:TFT:DensitySpaceHeadwayRelation}, there 
exists a relation between the flow $q$, and the average time headway $\overline 
h_{t}$ \cite{WARDROP:52,DAGANZO:97}:

\begin{equation}
\label{eq:TFT:FlowTimeHeadwayRelation}
	q = \frac{N}{T_{\text{mp}}}
	  = \frac{N}{\displaystyle \sum_{i = \one}^{N} h_{t_{i}}}
	  = \frac{\one}{\frac{\one}{N} \displaystyle \sum_{i = \one}^{N} h_{t_{i}}}
	  = \frac{\one}{\overline h_{t}},
\end{equation}

with ${\overline h_{t}}^{-\one}$ the reciprocal of the average time headway.

			\subsubsection{Oblique cumulative plots}
			\label{sec:TFT:ObliqueCumulativePlots}

\begin{figure}[!htb]
	\centering
	\includegraphics[width=\figurewidth]{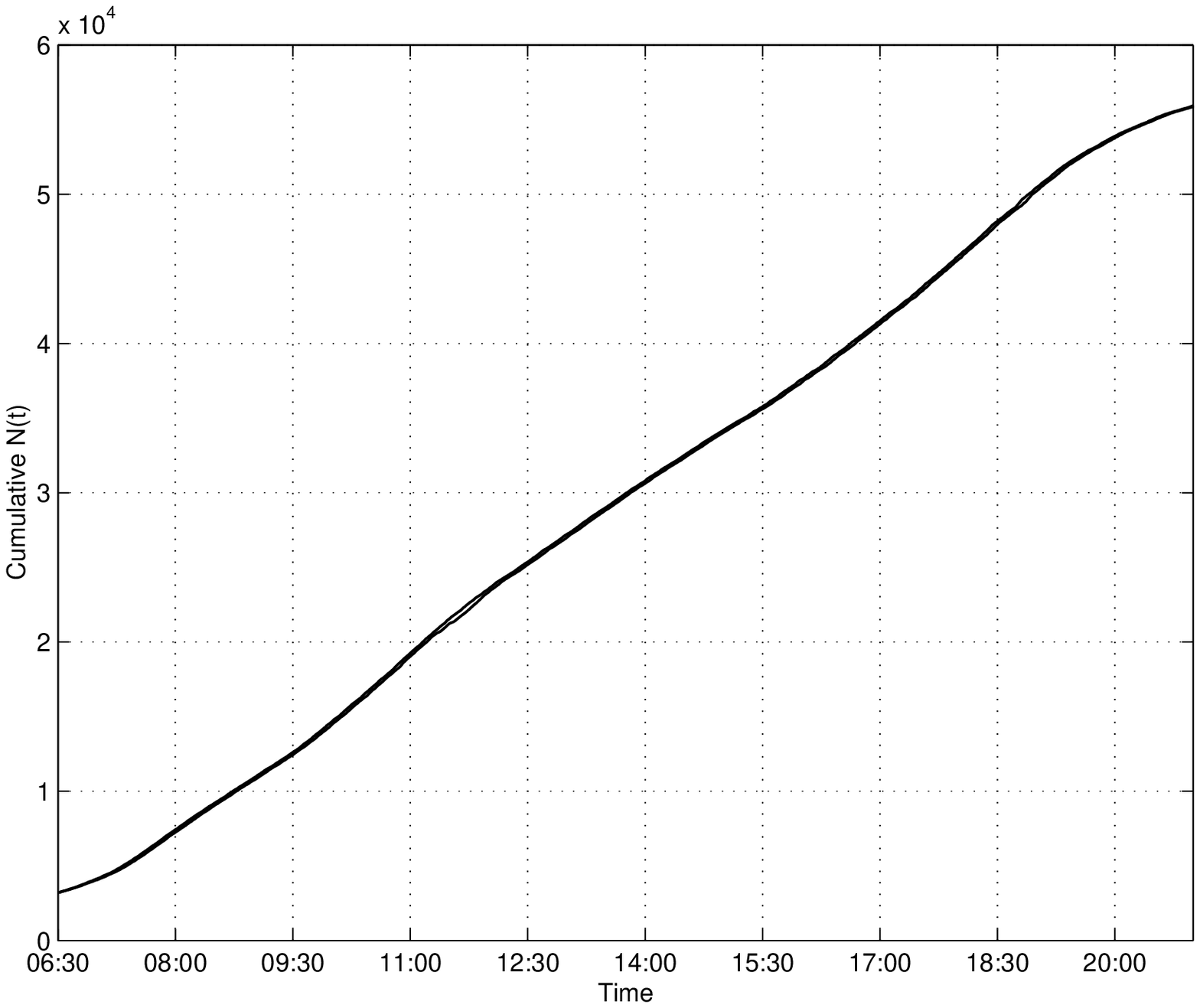}\\
	\vspace*{\figureseparation}
	\includegraphics[width=\figurewidth]{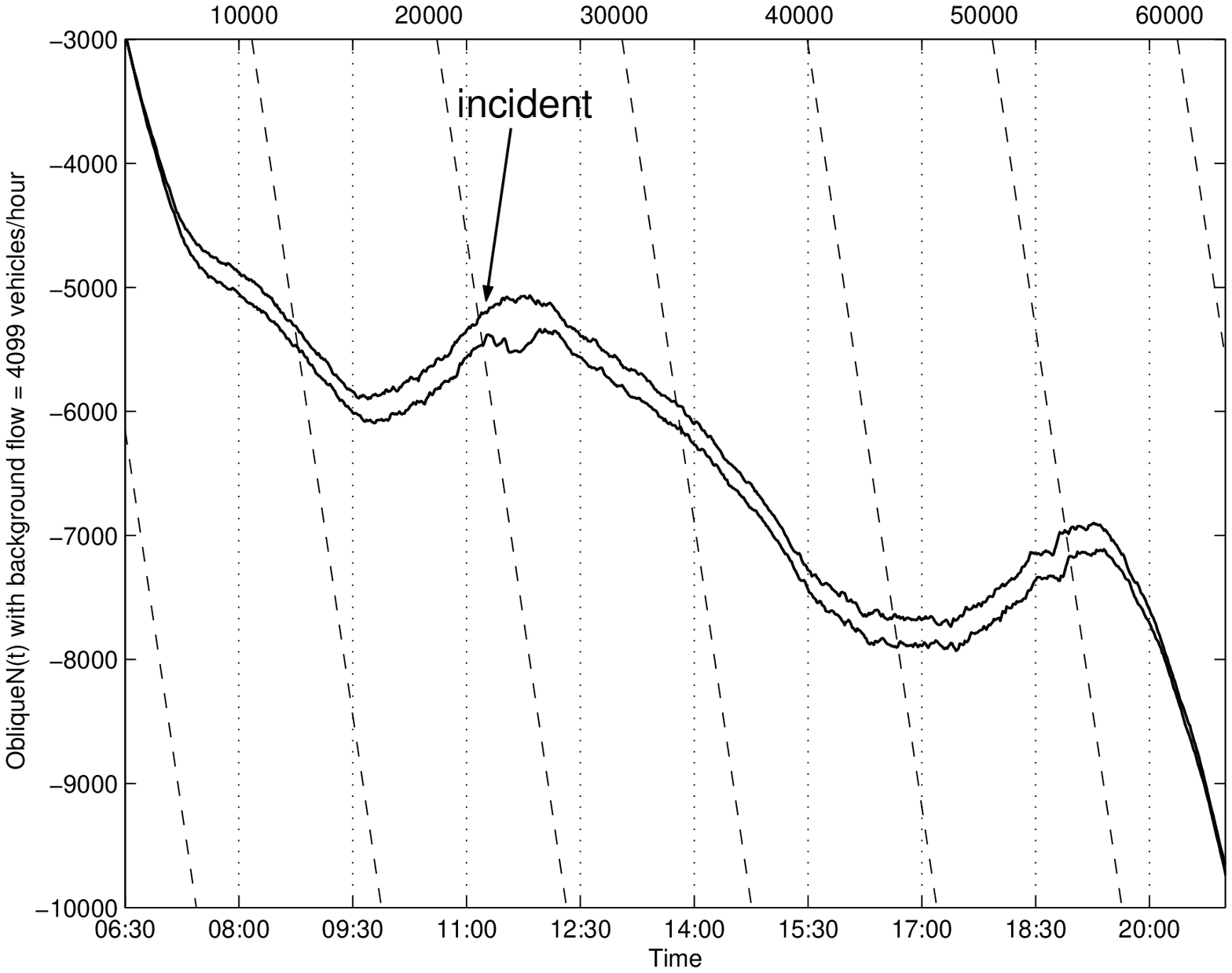}
	\caption{
		\emph{Left:} a standard cumulative plot showing the number of passing 
		vehicles at two detector locations; due to the graph's scale, both curves 
		appear to lie on top of each other. \emph{Right:} the same data but 
		displayed using an oblique coordinate system, thereby enhancing the 
		visibility (the dashed slanted lines have a slope corresponding to the 
		subtracted background flow $q_{b} \approx \text{4100 vehicles per hour}$). 
		We can see a queue (probably caused due to an incident) growing at 
		approximately 11:00, dissipating some time later at approximately 12:30. The 
		shown detector data was taken from single inductive loop detectors 
		\cite{VVC:03}, covering all three lanes of the E40 motorway between 
		Erpe-Mere and Wetteren, Belgium. The shown data was recorded at Monday April 
		4th, 2003 (the detectors' sampling interval was one minute, the distance 
		between the upstream and downstream detector stations was 8.1 kilometres).
	}
	\label{fig:TFT:ObliqueCumulativePlots}
\end{figure}

As stated before, flows are always expressed as a rate. In contrast to this, we 
can also consider the raw vehicle counts at a certain location (i.e., 
measurement region $R_{t}$). If we plot the cumulative number of passing 
vehicles (denoted by $N$) with respect to time for different regions (e.g., 
inductive loop detectors), we get a set of curves such as the one in the left 
part of \figref{fig:TFT:ObliqueCumulativePlots}. These curves are called 
\emph{cumulative plots} (or $(t,N)$ diagrams), and although their origins date 
back as far as 1954 with the work of Karl Moskowitz \cite{MOSKOWITZ:54}, it was 
Gordon Newell who applied them later on to their full potential (initially in 
the context of queueing theory) \cite{NEWELL:82,NEWELL:93,NEWELL:93b,NEWELL:93c} 
(a similar method was applied by John Luke, in the field of continuum mechanics 
\cite{LUKE:72,DAGANZO:95b}).

The key benefit of these cumulative plots, comes when comparing observations 
stemming from multiple detector stations at a closed section of the road that 
conserves the number of vehicles (i.e., no on- or off-ramps), in which case we 
also speak of \emph{input-output diagrams}. If there are two detector stations, 
then the upstream and downstream stations measure the \emph{input}, respectively 
\emph{output}, of the section. Similarly like in queueing theory, the upstream 
curve is sometimes called the \emph{arrival function}, whereas the downstream 
one is called the \emph{departure function} \cite{NEWELL:82}. As the method is 
based on counting the number of individual vehicles at each observation location 
(whereby each vehicle is numbered with respect to a single reference vehicle), 
this results in a monotonically increasing function $N(t)$ (sometimes called the 
\emph{Moskowitz function}, after its `inventor'), which increases each time a 
vehicles passes by. At each time instant $t$, the cumulative count is defined 
as:

\begin{equation}
	N(t) = \sum_{t' = t_{\zero}}^{t} q(t') = N(t - \one) + q(t).
\end{equation}

The time needed to travel from one location to another can easily be measured as 
the horizontal distance between the respective cumulative curves. Similarly, the 
vertical distance between these curves allows us to derive the accumulation of 
vehicles on the road section, which gives an excellent indication of growing and 
dissipating queues (i.e., congestion). Furthermore, if we compute the slope of 
this function at each time instant $t$, we obtain the flow $q(t) = [N(t + \Delta 
t) - N(t)] / \Delta t$. Finally, because $N(t)$ essentially is a step function, 
we can define a smooth approximation $\widetilde{N}(t)$. This results in an 
everywhere differentiable function, allowing us to compute instantaneous flows 
and local densities as $q = \partial \widetilde{N}(t,x) / \partial t$	and $k = - 
\partial \widetilde{N}(t,x) / \partial x$, respectively \cite{DAGANZO:97}.

The main disadvantage of the method is the fact that these cumulative functions 
increase very rapidly, thereby masking the subtle differences between different 
curves. Cassidy and Windover therefore proposed to subtract a background flow 
$q_{b}$ from these curves, resulting in functions $N(t) - t~q_{b}$ 
\cite{CASSIDY:95}. Based on this; Mu\~noz and Daganzo furthermore introduced 
enhanced clarity by overlaying this cumulative plot with a set of oblique lines 
with slope $-q_{b}$ \cite{MUNOZ:02}. Choosing an appropriate value for $q_{b}$, 
allows us to nicely enhance the characteristic undulations that are expressed by 
the different oblique curves.\\

\sidebar{
	Note that before using these oblique plots, the cumulative plots from 
	different detectors stations need to be \emph{synchronised}. To understand 
	this, suppose a reference vehicle passes an upstream detector station at a 
	certain time instant $t_{\text{up}}$; after a certain time period, the vehicle 
	reaches the downstream detector station at a later time instant 
	$t_{\text{down}}$. The amount $t_{\text{down}} - t_\text{up}$ is the time it 
	takes to cross the distance between both detector stations, allowing the 
	synchronisation mechanism to shift the respective cumulative curves over this 
	time period (i.e., initialising them with the passing of the reference 
	vehicle).\\

	One way to achieve this, is by looking at the respective shapes of both 
	cumulative curves during light traffic conditions (e.g., the early morning 
	period when free-flow conditions are prevailing). The idea now is to shift one 
	curve such that the difference between the two curves' shapes is minimal 
	\cite{WESTERMAN:95,MUNOZ:00,MUNOZ:03b}. Note that other corrections may be 
	necessary, as both detector stations can count a different number of vehicles 
	(i.e., a systematic bias).
}\\

An example of an oblique plot can be seen in the right part of 
\figref{fig:TFT:ObliqueCumulativePlots}: the cumulative count at each time 
instant can be read from an axis that is perpendicular to the oblique (slanted) 
overlayed dashed lines (e.g., we can see a count of some 30000 vehicles at 
14:00). Note that the accumulation can still be measured by the vertical 
distance between two curves (i.e., at a specific time instant), but the travel 
time should now be measured along one of the overlayed oblique lines. Such a 
pair of cumulative curves can be thought of as a flexible plastic garden hose: 
whenever there is an obstruction on the road, the outflow of the section will be 
blocked, resulting in a local thickening of this `hose' (i.e., the accumulation 
of vehicles on the section).

Using these oblique cumulative plots, we can now inspect the traffic dynamics in 
much more detail than was previously possible. For example, looking again at the 
right part of \figref{fig:TFT:ObliqueCumulativePlots}, we can see how the 
specific traffic stream characteristics propagate from one detector station to 
another. Even more visible, is a queue that starts to grow at approximately 
11:00 (i.e., the time of the appearance of a `bulge'), dissipating at 
approximately 12:30. As data curves from upstream detectors lie above data 
curves from downstream detectors, we see a decrease in the road section's 
output. Careful investigation of the traffic data revealed that the detector 
stations recorded a rather low flow (approximately 2500 vehicles per hour as 
opposed to a nominal flow of 4500 vehicles per hour), whereby all vehicles drove 
at a low speed (between 20 and 60 km/h as opposed to 110 km/h). This gives 
sufficient evidence to conclude that an incident probably occurred shortly after 
11:00, consequently obstructing a part of the road and leading to a build up of 
vehicles in the section.

Let us finally note that although oblique cumulative plots currently are not a 
mainstream technique used by the traffic community, we predict their rising 
popularity: they are one of the most simple, yet powerful, techniques for 
studying local traffic phenomena, giving traffic engineers practical insight 
into the formation of bottlenecks. Some recent examples include the work of 
Mu\~noz and Daganzo \cite{MUNOZ:00b,MUNOZ:00,MUNOZ:02b,MUNOZ:03b}, Cassidy and 
Bertini \cite{CASSIDY:99,BERTINI:03}, Cassidy and Mauch \cite{CASSIDY:01}, and 
Bertini et al. \cite{BERTINI:05}.

		\subsection{Occupancy}
		\label{sec:TFT:Occupancy}

Notwithstanding the importance of measuring traffic density, most of the 
existing detector stations on the road are only capable of temporal measurements 
(i.e., region $R_{t}$). If individual vehicle speeds can be measured, by double 
inductive loop detectors (DLD) for example, then density should be computed 
using equation \eqref{eq:TFT:DensityRegionTSingleLane}.

However, in many cases these vehicle speeds are not readily available, e.g., 
when using single inductive loop detectors. The detector's logic therefore 
resorts to a temporal measurement called the \emph{occupancy} $\rho$, which 
corresponds to the fraction of time the measurement location was occupied by a 
vehicle:

\begin{equation}
\label{eq:TFT:Occupancy}
	\rho = \frac{\one}{T_{\text{mp}}} \sum_{i = \one}^{N} o_{t_{i}}.
\end{equation}

In the previous equation, $o_{t_{i}}$ denotes the \ith vehicle's \emph{on-time}, 
i.e., the time period during which it is present above the detector (it 
corresponds to the shaded area swept by a vehicle at a certain location $x_{i}$ 
in \figref{fig:TFT:SpaceAndTimeHeadways}). Note that this on-time actually 
corresponds to the effective vehicle length as seen by the detector, divided by 
the vehicle's speed \cite{COIFMAN:01}:

\begin{equation}
	o_{t_{i}} = \frac{l_{i} + K_{\text{ld}}}{v_{i}},
\end{equation}

with $l_{i}$ the vehicle's true length and $K_{\text{ld}} > dx$ the finite, 
non-infinitesimal length of the detection zone. If we define $\overline o_{t}$ 
as the average on-time (based on the vehicles that have passed the detector 
during the observation period), then we can establish a relation between the 
occupancy and the flow \cite{DAGANZO:97} using equations 
\eqref{eq:TFT:FlowRegionTSingleLane} and \eqref{eq:TFT:Occupancy}:

\begin{equation}
	\rho = \left ( \frac{N}{T_{\text{mp}}} \right) ~ \left ( \frac{\one}{N} ~ \sum_{i = \one}^{N} o_{t_{i}} \right ) = q~\overline o_{t}.
\end{equation}

Furthermore, it is as before possible to define the occupancy for generalised 
measurement regions, using the total space consumed by the shaded areas of 
vehicles in a time-space diagram (e.g., \figref{fig:TFT:SpaceAndTimeHeadways}), 
divided by the area of the measurement region 
\cite{EDIE:65,DAGANZO:97,CASSIDY:98}. Continuing our discussion, assume that 
individual vehicle lengths and speeds are uncorrelated; it can then be shown 
that \cite{DAGANZO:97,COIFMAN:01}:

\begin{equation}
\label{eq:TFT:OccupancyDensity}
	\rho = \overline l~k \Longrightarrow k = \frac{\rho}{\overline l},
\end{equation}

with $\overline l$ the average vehicle length (note that this can correspond to 
the concept of passenger car units defined in section \ref{sec:TFT:Density}). 
Multiplying equation \eqref{eq:TFT:OccupancyDensity} by 100, allows us to 
express the occupancy as a percentage. For multi-lane traffic, the occupancy is 
derived in analogy to equation \eqref{eq:TFT:DensityRegionTMultiLane}:

\begin{equation}
	\rho = \sum_{i = \one}^{L} \rho_{l} = \frac{\one}{T_{\text{mp}}} \sum_{l = \one}^{L} \displaystyle \sum_{i = \one}^{N_{l}} o_{t_{i,l}},
\end{equation}

with now $o_{t_{i,l}}$ the on-time of the \ith vehicle in lane $l$. Note that 
the total occupancy derived in this way, can exceed 1 (but is bounded by $L$); 
if desired, it can be normalised through a division by $L$ to obtain the 
\emph{average occupancy}.

Note that if we apply equation \eqref{eq:TFT:OccupancyDensity} to measurement 
region $R_{s}$ based on the density in equation 
\eqref{eq:TFT:DensityRegionSSingleLane}, then the occupancy $\rho$ can be 
written as:

\begin{equation}
\label{eq:TFT:OccupancySingleLaneRegionRs}
	\rho = \left( \frac{\one}{\cancel{N}} \sum_{i = \one}^{N} l_{i} \right) \frac{\cancel{N}}{K} = \frac{\one}{K} \sum_{i = \one}^{N} l_{i}.
\end{equation}

So the occupancy now represents the `real density' of the road, i.e., the 
physical space that all vehicles occupy.\\

\sidebar{
	In the past, density was sometimes referred to as \emph{concentration}. 
	Nowadays however, concentration is used in a more broad context, encompassing 
	both density and occupancy whereby the former is meant to be a spatial 
	measurement, as opposed to the latter which is considered to be a temporal 
	measurement \cite{GARTNER:97}.
}\\

		\subsection{Mean speed}

The final macroscopic characteristic to be considered, is the \emph{mean speed} 
of a traffic stream; it is expressed in kilometres (or miles) per hour (the 
inverse of a vehicle's speed is called its \emph{pace}). Note that speed is not 
to be confused with \emph{velocity}; the latter is actually a vector, implying a 
direction, whereas the former could be regarded as the norm of this vector.

		\subsubsection{Mathematical formulation}

If we base our approach on direct measurements of the individual vehicles' 
speeds, we can generally obtain the mean speed as \emph{the total distance 
travelled by all the vehicles in the measurement region, divided by the total 
time spent in this region} \cite{EDIE:65,DAGANZO:97}. This gives the following 
derivations for the spatial and temporal regions, $R_{s}$ and $R_{t}$ 
respectively:

\begin{equation}
\label{eq:TFT:SpaceMeanSpeedSingleLane}
	\overline v_{s} = \frac{\displaystyle \sum_{i = \one}^{N} X_{i}}{\displaystyle \sum_{i = \one}^{N} T_{i}} =
		\left \lbrace
			\begin{array}{ll}
				\displaystyle \frac{\displaystyle \sum_{i = \one}^{N} v_{i}~\cancel{dt}}{\displaystyle N~\cancel{dt}} =
				\displaystyle \frac{\one}{N} \sum_{i = \one}^{N} v_{i} & \text{(region $R_{s}$)},\\
				& \\
				\displaystyle \frac{N~\cancel{dx}}{\displaystyle \sum_{i = \one}^{N} \frac{\cancel{dx}}{v_{i}}} =
				\displaystyle \frac{\one}{\displaystyle \frac{\one}{N} \displaystyle \sum_{i = \one}^{N} \frac{\one}{v_{i}}} & \text{(region $R_{t}$)},
			\end{array}
		\right.
\end{equation}

with now $X_{i}$ and $T_{i}$ the distance, respectively time, travelled by the 
\ith vehicle and $N$ the number of vehicles present during the measurement. The 
mean speed computed by the previous equations, is called the \emph{average 
travel speed} (the computation also includes stopped vehicles), which is more 
commonly known as the \emph{space-mean speed} (SMS); we denote it with 
$\overline v_{s}$ (note that in some engineering disciplines, the sole letter 
$u$ is used to denote a mean speed, however, this is ambiguous in our opinion).

It is interesting to see that the spatial measurement is based on an 
\emph{arithmetic average} of the vehicles' \emph{instantaneous speeds}, whereas 
the temporal measurement is based on the \emph{harmonic average} of the 
vehicles' \emph{spot speeds}. If we instead were to take the arithmetic average 
of the vehicles' spot speeds in the temporal measurement region $R_{t}$, this 
would lead to what is called the \emph{time-mean speed} (TMS); we denote it by 
$\overline v_{t}$:

\begin{equation}
\label{eq:TFT:TimeMeanSpeedSingleLane}
	\overline v_{t} = \frac{\one}{N} \sum_{i = \one}^{N} v_{i} \text{\quad (region $R_{t}$)}.
\end{equation}

Similarly, we can compute the time-mean speed for measurement region $R_{s}$, by 
taking the harmonic average of the vehicles' instantaneous speeds. With respect 
to both space- and time-mean speeds, Wardrop has shown that the following 
relation holds \cite{WARDROP:52}:

\begin{equation}
\label{eq:TFT:WardropSMSAndTMSRelation}
	\overline v_{t} = \overline v_{s} + \frac{\sigma_{s}^{\two}}{\overline v_{s}},
\end{equation}

with $\sigma_{s}^{\two}$ the statistical sample variance defined as follows:

\begin{equation}
	\sigma_{s}^{\two} = \frac{\one}{N - \one} \sum_{i = \one}^{N} (v_{i} - \overline v_{s})^{\two},
\end{equation}

in which $v_{i}$ denotes the \ith vehicle's instantaneous speed. One of the main 
consequences of equation \eqref{eq:TFT:WardropSMSAndTMSRelation}, is that the 
time-mean speed always exceeds the space-mean speed (except when all the 
vehicles' speeds are the same, in which case the sample variance is zero and, as 
a consequence, the time- and space-mean speeds are equal). So a stationary 
observer will most likely see more faster than slower vehicles passing by, as 
opposed to e.g., an aerial photograph in which more slower than faster vehicles 
will be seen \cite{DAGANZO:97}. Despite this mathematical quirk, the practical 
difference between SMS and TMS is often negligible for free-flow traffic (i.e., 
light traffic conditions); however, under congested traffic conditions both mean 
speeds will behave substantially differently (i.e., around 10\%).

Using equation \eqref{eq:TFT:WardropSMSAndTMSRelation}, we can also estimate the 
space-mean speed, based on the time-mean speed and approximating the variance of 
the SMS with that of the TMS \cite{BOVY:00}:

\begin{eqnarray}
	\overline v_{s} & = & \overline v_{t} - \frac{\sigma_{s}^{\two}}{\overline v_{s}},\nonumber\\
	                & \approx & \overline v_{t} - \frac{\sigma_{t}^{\two}}{\overline v_{s}},\nonumber\\
	                & \Downarrow & \nonumber\\
	\overline v_{s} - \overline v_{t} & \approx & - \frac{\sigma_{t}^{\two}}{\overline v_{s}},\nonumber\\
	\overline v_{s}^{\two} - \overline v_{s} \overline v_{t} & \approx & - \sigma_{t}^{\two},\nonumber\\
	\overline v_{s}^{\two} - \two~\overline v_{s} \frac{\overline v_{t}}{\two} + \frac{\overline v_{t}^{\two}}{\four} & \approx & \frac{\overline v_{t}^{\two}}{\four} - \sigma_{t}^{\two},\nonumber\\
	\left( \overline v_{s} - \frac{\overline v_{t}}{\two} \right)^{\two} & \approx & \frac{\overline v_{t}^{\two}}{\four} - \sigma_{t}^{\two},\nonumber\\
	                & \Downarrow & \nonumber\\
	\overline v_{s} & \approx & \frac{\overline v_{t}}{\two} + \sqrt{\frac{\overline v_{t}^{\two}}{\four} - \sigma_{t}^{\two}} \qquad \forall~\overline v_{t} \geq \two~\sigma_{t}.
\end{eqnarray}

In general, using the space-mean speed is preferred to the time-mean speed. 
However, in most cases only this latter traffic flow characteristic is 
available, so care should be taken when interpreting the results of a study 
(unless of course when SMS and TMS are negligibly different).

The extension of equation \eqref{eq:TFT:SpaceMeanSpeedSingleLane} to multi-lane 
is straightforward; for example, the space-mean speed is computed as follows:

\begin{equation}
\label{eq:TFT:SpaceMeanSpeedMultiLane}
	\overline v_{s} =
		\left \lbrace
			\begin{array}{ll}
				\displaystyle \sum_{l = \one}^{L} \sum_{i = \one}^{N_{l}} v_{i,l} ~ \left / ~ \sum_{l = \one}^{L} N_{l}\right. & \text{(region $R_{s}$)},\\
				& \\
				\displaystyle \frac{\one}{\displaystyle \frac{\one}{\displaystyle \sum_{l = \one}^{L} N_{l}} \displaystyle \sum_{l = \one}^{L} \sum_{i = \one}^{N_{l}} \frac{\one}{v_{i,l}}} & \text{(region $R_{t}$)},
			\end{array}
		\right.
\end{equation}

with now $v_{i,l}$ the instantaneous (or spot) speed of the \ith vehicle in lane 
$l$.

			\subsubsection{Fundamental relation of traffic flow theory}
			\label{sec:TFT:FundamentalRelation}

There exists a unique relation between three of the previously discussed 
macroscopic traffic flow characteristics density $k$, flow $q$, and space-mean 
speed $\overline v_{s}$ \cite{WARDROP:52}:

\begin{equation}
\label{eq:TFT:FundamentalRelation}
	q = k~\overline v_{s}.
\end{equation}

This relation is also called the \emph{fundamental relation of traffic flow 
theory}, as it provides a close bond between the three quantities: knowing two 
of them allows us to calculate the third one (note that the time-mean speed in 
equation \eqref{eq:TFT:TimeMeanSpeedSingleLane} does not obey this relation). In 
general however, there are two restrictions, i.e., the relation is only valid 
for (1) continuous variables\footnote{Note that the hypothesis also assumes that 
the variables are spatially measured, e.g., space-mean speed.}, or smooth 
approximations of them, and (2) traffic composed of substreams (e.g., slow and 
fast vehicles) which comply to the following two assumptions:

\begin{quote}
	\begin{description}
		\item[Homogeneous traffic]~

			There is a homogeneous composition of the traffic substream (i.e., the 
			same type of vehicles).
		\item[Stationary traffic]~

			When observing the traffic substream at different times and locations, it 
			`looks the same'. Putting it a bit more quantitatively, all the vehicles' 
			trajectories should be parallel and equidistant \cite{DAGANZO:97}. A 
			stationary time period can be seen in a cumulative plot (e.g., 
			\figref{fig:TFT:ObliqueCumulativePlots}) where the curve corresponds to a 
			linear function.
	\end{description}
\end{quote}

The latter of the above two conditions, is also referred to as traffic operating 
in a \emph{steady state} or at \emph{equilibrium}. Based on equations 
\eqref{eq:TFT:DensityRegionSMultiLane} and \eqref{eq:TFT:FlowRegionTMultiLane} 
using partial densities and flows for different substreams (e.g., vehicle 
classes with distinct travel speeds, macroscopic characteristics of different 
lanes, \ldots), we can now calculate the space-mean speed, using relation 
\eqref{eq:TFT:FundamentalRelation}, in the following equivalent ways:

\begin{eqnarray}
	\overline v_{s} & = & q~/~k,\nonumber\\
	                & = & \sum_{c = \one}^{C} q_{c} ~ \left / ~ \sum_{c = \one}^{C} k_{c},\right.\label{eq:TFT:SMSFromQK}\\
	                & = & \sum_{c = \one}^{C} q_{c} ~ \left / ~ \sum_{c = \one}^{C} \frac{q_{c}}{\overline v_{s_{c}}},\right.\label{eq:TFT:SMSFromQV}\\
	                & = & \sum_{c = \one}^{C} k_{c} ~ \overline v_{s_{c}} ~ \left / ~ \sum_{c = \one}^{C} k_{c},\right.\label{eq:TFT:SMSFromKV}
\end{eqnarray}

in which $C$ denotes the number of substreams, $q_{c}$, $k_{c}$, $\overline 
v_{s_{c}}$, and $\overline v_{t_{c}}$ the flow, density, space, and time-mean 
speed respectively of the $c$-th substream. In the above derivations, equation 
\eqref{eq:TFT:SMSFromQK} should be used when both the flows and densities are 
known, equation \eqref{eq:TFT:SMSFromQV} should be used when both the flows and 
space-mean speeds are known, and equation \eqref{eq:TFT:SMSFromKV} should be 
used when both the densities and space-mean speeds are known.

As can be seen in equation \eqref{eq:TFT:SMSFromKV}, the space-mean speed is 
calculated by averaging the substreams' space-mean speeds using their densities 
as weighting factors. Similarly, the time-mean speed can be derived by using the 
flows as weighting factors for the substreams' time-mean speeds:

\begin{equation}
	\overline v_{t} = \sum_{c = \one}^{C} q_{c} ~ \overline v_{t_{c}} ~ \left / ~ \sum_{c = \one}^{C} q_{c},\right.
\end{equation}

Because density can not always be easily measured, we can compute it using the 
fundamental relation \eqref{eq:TFT:FundamentalRelation}. Density can then be 
directly derived from flow and space-mean speed measurements, or if the latter 
are not available, they can be estimated from occupancy measurements; in 
\cite{COIFMAN:01,COIFMAN:03,COIFMAN:03b}, Coifman provides a nice set of 
techniques for dealing with these difficulties.

			\subsection{Moving observer method and floating car data}
			\label{sec:TFT:MovingObserverAndFCD}

When measuring and/or computing the macroscopic traffic flow characteristics in 
the previous sections, we always assumed a fixed measurement region. There 
exists however yet another method, based on what is called a \emph{moving 
observer} \cite{WARDROP:54}. The idea behind the technique is to have a vehicle 
drive in both directions of a traffic flow, each time recording the number of 
oncoming vehicles and the net number of vehicles it gets overtaken by, as well 
as the times necessary to complete the two trips. Note that the assumption of 
stationary traffic still has to hold, i.e., the round trip should be completed 
before traffic conditions change significantly.

Using this method, it is then possible to derive the flow and density of the 
traffic stream in the direction of interest \cite{GARTNER:97,DAGANZO:97}. 
However, the main disadvantage of this method is that, in order to obtain an 
acceptable level of accuracy on a road with a low flow, a very large number of 
trips are required \cite{WARDROP:54,GARTNER:97,MULLIGAN:02}.

One of the techniques that has entered the picture during the last decade, is 
the use of so-called \emph{floating cars} or \emph{probe vehicles}. They can be 
compared to the moving observer method, but in this case, the vehicles are 
equipped with GPS and GSM(C)/GPRS devices that determine their locations based 
on the USA's NAVSTAR-GPS (or Europe's planned GNSS Galileo), and transmit this 
information to some operator. Initially, this allows an agency, e.g., a parcel 
delivery service, to track its vehicles throughout a network, based on their 
locations. Nowadays, the technique has evolved, resulting in several completed 
field tests of which the main goal was to estimate the traffic conditions based 
on a small number of probe vehicles. During field measurements, floating cars 
can mimic several types of behaviour, most notably by travelling at the traffic 
flows' mean speed, or by trying to travel at the road's speed limit, or even by 
chasing another randomly selected vehicle from the traffic stream.

Some examples of studies and experiments with floating car data (FCD) are given 
in the following. Firstly, Fastenrath gives an overview of a telematic field 
trial (\emph{VEhicle Relayed Dynamic Information}, VERDI) that addresses issues 
such as economical, political, and technical constraints \cite{FASTENRATH}. 
Secondly, Westerman provides an overview of available techniques for obtaining 
real-time road traffic information, with the goal of controlling the traffic 
flows through telematics \cite{WESTERMAN:95}, and Wermuth et al. describe a 
`TeleTravel System' used for surveying individual travel behaviour 
\cite{WERMUTH:00}. Then, Taale et al. compare travel times from floating car data 
with measured travel times (using a fleet of sixty equipped vehicles driving 
around in Rotterdam, The Netherlands), concluding that they correspond 
reasonably well \cite{TAALE:00}. Next, Michler derives the minimum percentage of 
vehicles necessary, in order to estimate traffic stream characteristics for 
certain traffic patterns (e.g., free-flow and congested traffic) based on rigid 
statistical grounds \cite{MICHLER:01}, and Linauer and Leihs measure the travel 
time between points in a road network, based on a high number of users that 
submit a low number of GSM hand-over messages \cite{LINAUER:03}. In addition, 
Demir et al. accurately reconstruct link travel times during periods of traffic 
congestion, using only a very limited number of FCD-messages with a small number 
of users \cite{DEMIR:03}. A final, more regional, example is the founding of the 
government-supported Belgian `Telematics 
Cluster'\footnote{\texttt{http://www.telematicscluster.be}}, a platform for 
encouraging the use of telematics solutions for ITS. The initiative already 
includes some 57 members, stimulating the cooperation between users, 
telecommunication companies, and the automotive industry.

In conclusion, we can state that the use of probe vehicles provides an effective 
way to gather accurate current travel times in a road network, thereby allowing 
good up-to-date estimations of traffic conditions. The technique will continue 
to grow and evolve, already by introducing personalised traffic information to 
drivers, based on their location and the surrounding traffic conditions. This 
development is furthermore stimulated by the fact that GSM market penetration 
still rises above 70\% \cite{LINAUER:03}, and it is our belief this will also be 
the case for personal GPS devices and in-vehicle route planners.\\

\sidebar{
	Despite the obvious major advantage of obtaining accurate information on the 
	traffic conditions, the technique suffers from a jurisprudential battle, in 
	that there are many delicately privacy concerns involved with respect to the 
	mobile operator that wants to track individual people's units (not to mention 
	the monetary cost associated with the numerously induced communications).
}\\

	\section{Performance indicators}

After considering the previously mentioned macroscopic traffic flow 
characteristics, we now take a look at some popular performance indicators used 
by traffic engineers when assessing the quality of traffic operations. We 
concisely discuss the peak hour factor, the reliability of travel times, the 
levels of service, and a measure of efficiency of a road. For a more complete 
overview, we refer the reader to \cite{SHAW:03}.

		\subsection{Peak hour factor}

During high flow periods in the peak hour, a possible indicator for traffic flow 
fluctuations is the so-called \emph{peak hour factor} (PHF). It is calculated 
for one day as the average flow during the hour with the maximum flow, divided 
by the peak flow rate during one quarter hour within this hour \cite{MAY:90}:

\begin{equation}
\label{eq:TFT:PHF}
	\text{PHF} = \frac{\overline q_{|\text{60}}}{\overline q_{|\text{15}}}.
\end{equation}

For example, suppose we measure flows on a main unidirectional road with three 
lanes, during a morning peak: from 07:00 to 08:00 we measure consecutively 3500, 
6600, 6200, and 4500 vehicles/hour during each quarter. The total average flow 
$\overline q_{|\text{60}}$ is 5200 vehicles/hour, with a peak 15 minute flow 
rate $\overline q_{|\text{15}} = $~6600 vehicles/hour. The PHF is therefore 
equal to $\text{5200} / \text{6600} = \text{0}.\overline{\text{78}}$.

Note that some manuals express the peak 15 minute flow rate as the number of 
vehicles during that quarter hour, necessitating an extra multiplication by 4 in 
the denominator of equation \eqref{eq:TFT:PHF} to convert the flow rate to an 
hourly rate.

We can immediately see that the PHF is constrained to the interval [0.25,1.00]; 
the higher the PHF, the flatter the peak period (i.e., a longer sustained state 
of high flow). Typically, the PHF has values around 0.7 -- 0.98. Note that two 
of the obvious problems with the PHF are, on the one hand, the question of when 
to pick the correct 15 minute interval, and on the other hand the fact that some 
peak periods may last longer than one hour.

	\subsection{Travel times and their reliability}

When travelling around, people like to know how long a specific journey will 
take (e.g., by public transport, car, bicycle, \ldots). This notion of an 
expected travel time, is one of the most tangible aspects of journeying as 
perceived by the travellers. When people are travelling to their work, they are 
required to arrive on time at their destinations. Based on this premise, we can 
naturally state that people reason with a built-in safety margin: they consider 
the \emph{average time} it takes to reach a destination, and use this to decide 
about their departure time.

Aside from the above obvious human rationale, there is also an increased 
interest in obtaining precise information with respect to travel times in the 
context of \emph{advanced traveller information systems} (ATIS). Here, an 
essential ingredient is the accurate prediction of future travel times. Coupled 
with incident detection for example, drivers can obtain correct travel time 
information, thereby staying informed of the actual traffic conditions and 
possibly changing their journey. The requested information can reach the driver 
by means of a cell-phone (e.g., as a feature offered by the mobile service 
provider), it can be broadcasted over radio (e.g., the \emph{Traffic Message 
Channel} -- TMC), or it can be displayed using \emph{variable message signs} 
(VMS) above certain road sections (e.g., \emph{dynamic route information panels} 
-- DRIPs), \ldots

		\subsubsection{Travel time definitions}

The travel time of a driver completing a journey, can be defined as `the time 
necessary to traverse a route between any two points of interest' 
\cite{TURNER:98}. In this context, the \emph{experienced dynamic travel time}, 
starting at a certain time $t_{\zero}$, over a road section of length $K$ is 
defined as follows \cite{BOVY:00}:

\begin{equation}
	T(t_{\zero}) = \int_{\zero}^{K} \frac{\one}{v(t,x)} dx \qquad \forall~t \geq t_{\zero},
\end{equation}

for which it is assumed that all local \emph{instantaneous} vehicle speeds 
$v(t,x)$ are known at all points along the route, and at all time instants 
(hence the term \emph{dynamic} travel time). In most cases however, we do not 
know all the $v(t,x)$, but only a finite subset of them, defined by the 
locations of the detector stations (demarcating section boundaries). The travel 
time can then be approximated using the recorded speeds at the beginning and end 
of a section (there is an underlying assumption here, namely that vehicles 
travel at a more or less constant speed between detector locations). As stated 
earlier, the experienced travel time requires the knowledge of local vehicle 
speeds at \emph{all} time instants after $T_{\zero}$. Because this is not always 
possible, a simplification can be used, resulting in the so-called 
\emph{experienced instantaneous travel time}:

\begin{equation}
	\widetilde{T}(t_{\zero}) = \int_{\zero}^{K} \frac{\one}{v(t_{\zero},x)} dx,
\end{equation}

In general, we can derive the travel time using equation 
\eqref{eq:TFT:SpaceMeanSpeedSingleLane}, i.e., the total distance travelled by 
all the vehicles, divided by their space-mean speed:

\begin{equation}
\label{eq:TFT:TravelTimeAndSMS}
	T(t_{\zero}) = \frac{K}{\overline v_{s}(t_{\zero})},
\end{equation}

in which an accurate estimation of the space-mean speed $\overline 
v_{s}(t_{\zero})$ at time $t_{\zero}$ is necessary (e.g., by taking the harmonic 
average of the recorded spot speeds).

		\subsubsection{Queueing delays}

Traffic congestion nearly always leads to the build up of queues, introducing an 
increase (i.e., the \emph{delay}) in the experienced travel time. The congestion 
itself can have originated due to traffic demand exceeding the capacity, or 
because an incident occurred (e.g., road works, a traffic accident, 
\ldots)\footnote{Note that in a broader sense, queueing delays also encompass 
delays at signallised and unsignallised intersections.}. This can create 
\emph{incidental} (non-recurrent) or \emph{structural} (recurrent) congestion. 
Congestion can thus be seen as a loss in travel time with respect to some base 
line reference. Two such commonly used references are the travel time under 
free-flow conditions, and the travel time under maximum (i.e., capacity) flow. 
The delay is typically expressed in vehicle hours. As stated earlier, there are 
several ways to inform a driver of the current and predicted travel time. Using 
DRIPs it is possible to advertise the extra travel time (the delay is now 
typically expressed in vehicle minutes), as well as queue lengths. We note that 
in our opinion it is more intuitive to advertise a temporal estimation (i.e., 
the travel time or the delay), than a spatial estimation (e.g., the queue length 
on a motorway).

		\subsubsection{An example of travel time estimation using cumulative plots}

There exist several techniques for estimating the current travel time; one 
method for directly `measuring' the travel time, is by using a probe vehicle (we 
refer the reader to section \ref{sec:TFT:MovingObserverAndFCD} for more 
details). This way, it is possible to extract actual travel times from a traffic 
stream. Note that as traffic conditions get more congested, more probe vehicles 
are required in order to obtain an accurate estimation of the travel time.

Another method for measuring the travel time, is based on historical data, 
namely cumulative plots (introduced in section 
\ref{sec:TFT:ObliqueCumulativePlots}). As mentioned earlier, the travel time can 
then be measured as the distance along the horizontal (or oblique) time axis; 
any excess due to delays can then easily be spotted on a set of oblique 
cumulative plots. 

\begin{figure}[!htb]
	\centering
	\includegraphics[width=\figurewidth]{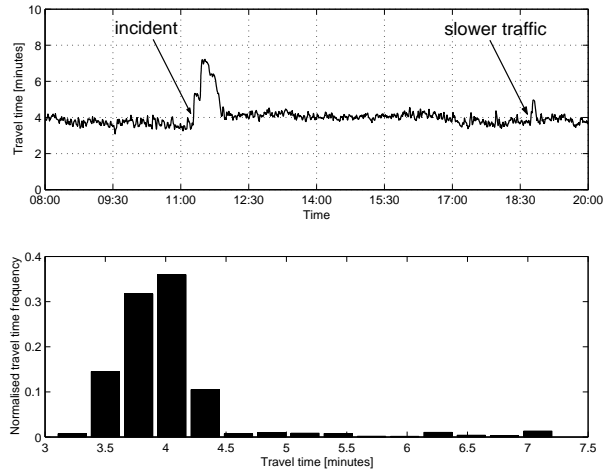}
	\caption{
		\emph{Top:} The evolution of the travel time during one day, based on the 
		cumulative plots from section \ref{sec:TFT:ObliqueCumulativePlots}. As can 
		be seen, an incident likely occurred at 11:00, increasing the travel time 
		from 4 to 7 minutes. Furthermore, at approximately 18:45 in the evening, all 
		traffic seemed to simultaneously slow down for a period of some 10 minutes. 
		\emph{Bottom:} Based on the calculated travel times during the day, we can 
		derive a histogram that is an approximation of the underlying travel time 
		probability density function. The mean is located around 4 minutes.
	}
	\label{fig:TFT:TravelTimeHistogram}
\end{figure}

Based on cumulative plots of consecutive detector stations, we can calculate the 
travel time between the upstream and downstream end of a road section. To 
illustrate this, let us reconsider the cumulative curves shown in 
\figref{fig:TFT:ObliqueCumulativePlots} of section 
\ref{sec:TFT:ObliqueCumulativePlots}. The evolution of the travel time during 
the day for these curves, is depicted in the top part of 
\figref{fig:TFT:TravelTimeHistogram}. The derived histogram (indicative of the 
underlying travel time probability density function), in the bottom part of the 
figure, shows that the mean travel time during the day is approximately 4 
minutes.

We already mentioned the likely occurrence of an incident at 11:00, resulting in 
the formation of a queue. During this period, the travel time shot up, reaching 
first 5, then 7 minutes. Looking at the top part of 
\figref{fig:TFT:TravelTimeHistogram}, we furthermore notice a slight increase in 
the travel time at approximately 18:45, for a short period of some 10 minutes. 
Investigation of the detector data, revealed that the flow remained constant at 
about 4500 vehicles per hour, but the speed dropped to some 90 km/h (as opposed 
to 110 km/h); we can conclude that all vehicles were probably simultaneously 
slowing down during this period (perhaps a rubbernecking effect). Another 
possibility is a platoon of slower moving vehicles, but then it would seem to 
have dissipated rather quickly after 10 minutes.

Using ample historical data, we can analyse the travel time over a period of 
many weeks, months, or even years. This would allow us to make \emph{intuitive} 
statements such as:

\begin{quote}
	\emph{``The typical travel time over this section of the road during a working 
	Monday, lies approximately between 4 and 6 minutes. There is however an 8\% 
	probability that the travel time increases to some 22 minutes (e.g., due to an 
	occurring incident).''}
\end{quote}

Finally note that, besides the two previously mentioned techniques for 
estimating travel times, an extensive overview can be found in the \emph{Travel 
Time Data Collection Handbook} \cite{TURNER:98}. Another concise but more 
theoretically-oriented overview is provided by Bovy and Thijs \cite{BOVY:00}.

		\subsubsection{Reliability and robustness properties}

As mentioned in the introduction of this section, people reason about their 
expected travel times based on a built-in safety margin. Central to this is the 
concept of the \emph{average travel time}. The \emph{reliability} of such a 
travel time is then characterised by its \emph{standard deviation}. Drivers 
typically accept (and sometimes expect) a small delay in their expected travel 
time. A traveller knows the \emph{expected travel time} because of the 
familiarity with the associated trip. To the traveller, this is personal 
historical information, for instance obtained by learning the trip's details 
(e.g., the traffic conditions during a typical morning rush hour) 
\cite{BALMER:04}.

Directly linked to the reliability of a certain expected travel time, is its 
variability. They are said to be unreliable when both expected and experienced 
travel times differ sufficiently. A typical characterisation of reliability 
involves the mean and standard deviation (i.e., the variance, which is a measure 
of variability) of a travel time distribution \cite{CHEN:02}. An example of such 
a travel time distribution for one day is shown in the histogram in the bottom 
part of \figref{fig:TFT:TravelTimeHistogram}.

Both first- and second-order measures of a distribution are by themselves 
insufficient to capture the complete picture. In order to grasp the notion of 
the previously mentioned safety margin, another typical statistical measure is 
considered, namely the \emph{90\theth percentile}. The rationale behind the use 
of this percentile is that travellers adopt a certain `safe' threshold with 
respect to their expected journey times. Considering the 90\theth percentile, 
this means that only one out of ten times the experienced travel time will 
differ significantly from the expected travel time. Travel time reliability can 
thus be viewed upon as a measure of service quality (similar to the concept of 
`quality of service' (QoS) in telecommunications).

There has been some research into the analytic form of travel time distributions 
(e.g., the work of Arroyo and Kornhauser, concluding that a lognormal 
distribution seems the most appropriate \cite{ARROYO:05}). There exist however 
significant differences between travel time distributions: in general, a smaller 
standard deviation indicates a better service quality and reliability. In 
contrast to this, a large standard deviation is indicative of chaotic behaviour 
of the traffic flow, the latter being totally unstable. Furthermore, travel time 
distributions can have a long tail; this signifies seldom events (e.g., 
incidents), that can have significant repercussions on the quality of traffic 
operations.\\

\sidebar{
	Let us finally note that there is an increased interest in the 
	\emph{reliability of complete transportation networks}, and their 
	\emph{robustness} against incidents. To this end, Immers et al. consider 
	reliability as a user-oriented quality, whereas robustness is more a property 
	of the system itself \cite{IMMERS:04}. Among several characterising factors 
	for robustness of transportation systems, they also introduce the following 
	practical notions in this context: \emph{redundancy}, denoting a spare 
	capacity, and \emph{resilience}, which is the ability to repeatedly recover 
	from a temporary overload. Their conclusion is that the key element in 
	securing transportation reliability lies in a good network design.
}\\

		\subsection{Level of service}
		\label{sec:TFT:LevelOfService}

Historically, one of the main performance indicators to assess the quality of 
traffic operations, was the \emph{level of service} (LOS), introduced in the 
1960s. It is represented as a grading system using one of six letters (A -- F), 
whereby LOS A denotes the best operating conditions and LOS F the worst. These 
LOS measures are based on road characteristics such as speed, travel time, 
\ldots, and drivers' perceptions of comfort, convenience, \ldots \cite{HCM:00}. 
As is customary among traffic engineers, these representative statistics of 
these characteristics are collectively called \emph{measures of effectiveness} 
(MOE).

Levels A through D are representative for free-flow conditions whereby LOS A 
corresponds to free flow, LOS B to reasonable free flow, LOS C to stable traffic 
operations, and LOS D to bordering unstable traffic operations. LOS E is 
reminiscent of near-capacity flow conditions that are extremely unstable, 
whereas LOS F corresponds to congested flow conditions (caused by either 
structural or incidental congestion) \cite{MAY:90}.

As an example, we provide an overview of the different levels of service in 
\tableref{table:TFT:LevelOfService} (based on \cite{MAY:90}, in similar form 
originally published in the \emph{Highway Capacity Manual} (HCM) of 1985 as the 
\emph{Transportation Research Board's} (TRB)\footnote{The TRB was formerly known 
as the \emph{Highway Research Board} (HRB).} special report \#209.

\begin{table}[!htb]
	\centering
	\begin{tabular}{|c|c|c|c|}
		\hline
		\emph{LOS} & \emph{Density (veh/km)} & \emph{Occupancy (\%)} & \emph{Speed (km/h)} \\
		\hline
		\hline 
		A & 0  $\rightarrow$ 7  & 0  $\rightarrow$ 5  & $\geq$~97\\
		\hline
		B & 7  $\rightarrow$ 12 & 5  $\rightarrow$ 8  & $\geq$~92\\
		\hline
		C & 12 $\rightarrow$ 19 & 8  $\rightarrow$ 12 & $\geq$~87\\
		\hline
		D & 19 $\rightarrow$ 26 & 12 $\rightarrow$ 17 & $\geq$~74\\
		\hline
		E & 26 $\rightarrow$ 42 & 17 $\rightarrow$ 28 & $\geq$~48\\
		\hline
		F & 42 $\rightarrow$ 62 & 28 $\rightarrow$ 42 & $<$~48\\
		  &    $>$ 62           &              $>$ 42 & \\
		\hline
	\end{tabular}
	\caption{
		Level of service (LOS) indicators for a motorway (adapted from 
		\cite{MAY:90}, in similar form originally published in the 1985 HCM).
	}
	\label{table:TFT:LevelOfService}
\end{table}

Calculating levels of service can be done using a multitude of methods; some 
examples include using the density (at motorways), using the space-mean speed 
(at arterial streets), using the delay (at signallised and unsignallised 
intersections), \ldots \cite{HCM:00}. The distinction between different LOS is 
primarily based on the measured average speed, and secondly on the density (or 
occupancy). Furthermore, as traditional analyses only focus on a select number 
of hours, a new trend is to conduct \emph{whole year analyses} (WYA) based on 
aggregated measurements such as e.g., the \emph{monthly average daily traffic} 
(MADT) and the \emph{annual average daily traffic} (AADT) \cite{BRILON:00}. The 
MADT is calculated as the average amount of traffic recorded during each day of 
the week, averaged over all days within a month. Averaging the resulting twelve 
MADTs gives the AADT.\\

\sidebar{
	Regarding the use of the LOS, we note that it is a rather old-fashioned method 
	for evaluating the quality of traffic operations. In general, it is difficult 
	to calculate, mainly because the defined standards at which the different 
	levels are set, always depend on the specific type of traffic situation that 
	is studied (e.g., type of road, \ldots). This makes the LOS more of an 
	engineering tool, used when assessing and planning operational analyses. 
	Instead of using the LOS, we therefore propose to adopt the more suited 
	approach based on oblique cumulative plots (we refer the reader to section 
	\ref{sec:TFT:ObliqueCumulativePlots}). These allow for example to assess the 
	differences between travel times under free-flow and congested conditions, 
	thereby giving a more meaningful and intuitive indication of the quality of 
	traffic operations to the drivers.
}\\

		\subsection{Efficiency}

In \cite{CHEN:01}, Chen et al. state that the main reason for congestion is not 
demand exceeding capacity (i.e., the number of travellers who \emph{want} to use 
a certain part of the transportation network, exceeds the available 
infrastructure's capacity), but is in fact the inefficient operation of 
motorways during periods of high demand. In order to quantify this efficiency, 
they first look at what the prevailing speed is when a motorway is operating at 
its maximum efficiency, i.e., the highest flow (corresponding to the effective 
capacity, which is different from the HCM's capacity which is calculated from 
the road's physical characteristics). Based on the distribution of 5-minute data 
samples from some 3300 detectors, they investigate the speed during periods of 
very high flows. This leads them to a so-called sustained speed $\overline 
v_{\text{sust}} = $~60 miles per hour (which corresponds to $\text{60~mi/h} 
\times \text{1.609} \approx \text{97}$~km/h).

The performance indicator they propose, is called the \emph{efficiency} $\eta$ 
and it based on the ratio of the \emph{total vehicle miles travelled} (VMT), 
divided by the \emph{total vehicle hours travelled} (VHT). Note that as the 
units of VMT and $v_{\text{sust}}$ should correspond to each other, we propose 
to use the terminology of \emph{total vehicle distance travelled} (VDT) instead 
of the VMT, in order to eliminate possible confusion. Both VDT and VHT are 
defined as follows:

\begin{eqnarray}
	\text{VDT} & = & q~K,\\
	           &   & \nonumber\\
	\text{VHT} & = & \frac{\text{VDT}}{\overline v_{s}},
\end{eqnarray}

with, as before, $q$ the flow, $K$ the length of the road section, and 
$\overline v_{s}$ the space-mean speed. Using the above definitions, we can 
write the efficiency of a road section as:

\begin{equation}
	\eta = \frac{\text{VDT} / \text{VHT}}{\overline v_{\text{sust}}}.
\end{equation}

The efficiency is expressed as a percentage, and it can rise above 100\% when 
the recorded average speeds surpass the sustained speed $\overline 
v_{\text{sust}}$. In general, the discussed efficiency can also easily be 
calculated for a complete road network and an arbitrary time period. It can 
furthermore be seen as the ratio of the actual productivity of a road section 
(the \emph{output} produced by this section during one hour), to its maximum 
possible production (the \emph{input} to the section) under high flow 
conditions.\\

\sidebar{
	Note that as a solution to their original claim (\emph{``congestion arises due 
	to inefficient operation''}), Chen et al. propose to increase the operational 
	efficiency, mainly through the technique of suitable ramp metering (using an 
	idealised ramp metering control practice that maintains the occupancy 
	downstream of an on ramp to its critical level). But in our opinion, they 
	neglect to take into account the entire situation, i.e., they fail to consider 
	the extra effects induced by holding vehicles back at some on ramps (e.g., the 
	total time travelled by \emph{all} the vehicles, including delays), thus 
	rendering their statement practically worthless by giving a feeble argument. 
	Careful examination of their reasoning, reveals that these extra effects are 
	dealt with by shifting demand during the peak periods\ldots but this just 
	confirms our hypothesis that congestion occurs when demand exceeds capacity, 
	even when this capacity is for example controlled through ramp metering~!
}\\

In contrast to the work of Chen et al., Brilon proposes another definition for 
the \emph{efficiency} (now denoted as $E$): it is expressed as the number of 
vehicle kilometres that are produced by a motorway section per unit of time 
\cite{BRILON:00}:

\begin{equation}
	E = q~\overline v_{s}~T_{\text{mp}},
\end{equation}

with now $q$ the total flow recorded during the time interval $T_{\text{mp}}$. 
Brilon concludes that in order for motorways to operate at maximum efficiency, 
their hourly flows typically have to remain \emph{below} the capacity flow 
(e.g., at 90\% of $q_{\text{cap}}$). Brilon also proposes to use this point of 
maximum efficiency as the threshold when going from LOS D to LOS E.

	\section{Fundamental diagrams}
	\label{sec:TFT:FundamentalDiagrams}

Whereas the previous sections dealt with individual traffic flow 
characteristics, this section discusses some of the relations between them. We 
first give some characterisations of different traffic flow conditions and the 
rudimentary transitions between them, followed by a discussion of the relations 
(which are expressed as fundamental diagrams) between the traffic flow 
characteristics, giving special attention to the different points of view 
adopted by traffic engineers.

		\subsection{Traffic flow regimes}
		\label{sec:TFT:TrafficFlowRegimes}

Considering a stream of traffic flow, we can distinguish different types of 
operational characteristics, called \emph{regimes} (two other commonly used 
terms are traffic flow \emph{phases} and \emph{states}). As each of these 
regimes is characterised by a certain set of unique properties, classification 
of them is sometimes based on occupancy measurements (see for example the 
discussion about levels of service in section \ref{sec:TFT:LevelOfService}), or 
it is based on combinations of different macroscopic traffic flow 
characteristics (e.g., the work of Kerner \cite{KERNER:04}).

In the following sections, we discuss the regimes known as \emph{free-flow} 
traffic, \emph{capacity-flow} traffic, \emph{congested}, \emph{stop-and-go}, and 
\emph{jammed} traffic. Our discussion of these regimes is in fact based on the 
commonly adopted way of looking at traffic flows, as opposed to for example 
Kerner's three-phase traffic theory that includes a regime known as 
\emph{synchronised} traffic (we refer the reader to section 
\ref{sec:TFT:Kerners3PT} for more details). We conclude the section with a note 
on the transitions that occur from one regime to another.

			\subsubsection{Free-flow traffic}
			\label{sec:TFT:FreeFlowTraffic}

Under light traffic conditions, vehicles are able to freely travel at their 
desired speed. As they are largely unimpeded by other vehicles, drivers strive 
to attain their own comfortable travelling speed (we assume that in case a 
vehicle encounters a slower moving vehicle ahead, it can easily change lanes in 
order to overtake the slower vehicle). Notwithstanding this ability for 
unconstrained travelling, drivers have to take into account the maximum allowed 
speed (denoted by $v_{\text{max}}$), as well as road-, engine-, and other 
vehicle characteristics. Note that in some cases, depending on the country under 
scrutiny, drivers perform speeding.

In essence, the previous description of \emph{free-flow traffic} considers a 
traffic flow to be unrestricted, i.e., no significant delays are introduced due 
to possible overtaking manoeuvres. As a consequence, the \emph{free-flow speed} 
(by some called the \emph{nominal speed}) is the mean speed of all vehicles, 
travelling at their own pace (e.g., 100~km/h); it is denoted by $\overline 
v_{\text{ff}}$.

Free-flow traffic occurs exclusively at low densities, implying large average 
space headways according to equation \eqref{eq:TFT:DensitySpaceHeadwayRelation}. 
As a result, small local disturbances in the temporal and spatial patterns of 
the traffic stream have no significant effects, hence traffic flow is 
\emph{stable} in the free-flow regime.

			\subsubsection{Capacity-flow traffic}
			\label{sec:TFT:CapacityFlowTraffic}

When the traffic density increases, vehicles are driving closer to each other. 
Considering the number of vehicles that pass a certain location alongside the 
road, an observer will notice an increase in the flow. At a certain moment, the 
flow will reach a maximum value (which is determined by the mean speed of the 
traffic stream and the current density). This maximum flow is called the 
\emph{capacity flow}, denoted by $q_{c}$, $q_{\text{cap}}$, or even 
$q_{\text{max}}$. A typical value for the capacity flow on a three-lane Belgian 
motorway with $v_{\text{max}}$ equal to 120~km/h, can reach a maximum of some 
7000 vehicles \cite{VVC:03}. According to equation 
\eqref{eq:TFT:FlowTimeHeadwayRelation}, the average time headway is minimal at 
capacity-flow traffic, indicating the (local) formation of tightly packed 
clusters of vehicles (i.e., platoons), which are moving at a certain 
\emph{capacity-flow speed} $\overline v_{c}$ (or $\overline v_{\text{cap}}$) 
which is normally a bit lower than the free-flow speed. Note that some of these 
fast platoons are very unstable when they are composed of tail-gating vehicles: 
whenever in such a string a vehicle slows down a little, it can have a cascading 
effect, leading to exaggerate braking of following vehicles. Hence, these latter 
manoeuvres can destroy the local state of capacity-flow, and can in the worst 
case lead to multiple rear-end collisions. At this point, traffic becomes 
\emph{unstable}.\\

\sidebar{
	The calculation of the capacity flow is a daunting task, holding traffic 
	engineers occupied for the last six decades. The fact of the matter is that 
	there exists no rigourous definition for the concept of `capacity'. As a result, 
	after many years of research, this culminated in the publication of the fourth 
	edition of the already previously mentioned \emph{Highway Capacity Manual}. It 
	contains an impressive overview, spanning methodologies for assessing the 
	capacity at specific types of road infrastructures (motorway facilities, weaving 
	sections, on- and off-ramps, signallised and unsignallised urban intersections, 
	\ldots) \cite{HCM:00}.
}\\

			\subsubsection{Congested, stop-and-go, and jammed traffic}
			\label{sec:TFT:CongestedStopAndGoAndJammedTraffic}

Considering the regime of capacity-flow traffic, it is reasonable to assume that 
drivers are more mentally aware and alert in this regime, as they have to adapt 
their driving style to the smaller space and time headways under high speeds. 
However, when more vehicles are present, the density is increased even further, 
allowing a sufficiently large disturbance to take place. For example, a driver 
with too small space and time headways, will have to brake in order to avoid a 
collision with the leader directly in front; this can lead to a local chain of 
reactions that disrupts the traffic stream and triggers a breakdown of the flow. 
The resulting state of saturated traffic conditions, is called \emph{congested 
traffic}. The moderately high density at which this breakdown occurs, is called 
the \emph{critical density}, and is denoted by $k_{c}$ or $k_{\text{crit}}$ (for 
a typical motorway, its value lies around 25 vehicles (PCUs) per kilometre per 
lane, \cite{VVC:03}). From this knowledge, we can derive the optimal driving 
speed for single-lane traffic flows as $\overline v_{s} = q_{\text{cap}} / 
k_{\text{crit}} = \text{2000} \div \text{25} = \approx \text{85}$~km/h.

Higher values for the density indicate almost always a worsening of the traffic 
conditions; congested traffic can result in \emph{stop-and-go traffic}, whereby 
vehicles encounter so-called \emph{stop-and-go waves}. These waves require them 
to slow down severely, or even stop completely. When traffic becomes motionless, 
the space headway reaches a minimum as all vehicles are standing 
bumper-to-bumper; this extreme state is called \emph{jammed traffic}. Clearly, 
there exists a maximum density at which the traffic seems to turn into a 
`parking lot', called the \emph{jam density} and it is denoted by $k_{j}$, 
$k_{\text{jam}}$, or $k_{\text{max}}$. For a typical motorway, its value lies 
around 140 vehicles (PCUs) per kilometre per lane \cite{VVC:03}.\\

\sidebar{
	Note that the jam density is typically expressed in vehicles per kilometre. As 
	already stated in the introduction of section \ref{sec:TFT:Density}, density 
	ignores the effects of traffic composition and vehicle lengths. For a typical 
	value of some 140 vehicles/km/lane for the jam density, this means that we 
	express the density by using passenger car units (see section 
	\ref{sec:TFT:PCUs} for more details). Suppose now for example that an average 
	trailer truck equals 4.5 PCUs, then the jam density would decrease to some 
	140~$\div$~4.5~$\approx$~31 trucks for this class of vehicles. \emph{As a 
	consequence, the value of the jam density is different for each vehicle 
	class.}
}\\

			\subsubsection{A note on the transitions between different regimes}
			\label{sec:TFT:PhaseTransitions}

Streams of traffic flows can be regarded as many-particle systems (e.g., gasses, 
magnetic spin systems, \ldots); as they have a large number of degrees of 
freedom, it is often intractable when it comes to solving them exactly. However, 
from a physical point of view, these systems can be described in the framework 
of \emph{statistical physics}, whereby the collective behaviour of their 
constituents is approximately treated using statistical techniques.

Within this context, the changeover from one traffic regime to another, can be 
looked upon as a \emph{phase transition}. Within thermodynamics and statistical 
physics, an \emph{order parameter} is often used to describe the phase 
transition: when the system shifts from one phase to another (e.g., at a 
\emph{critical point} for liquid-gas transitions), the order parameter expresses 
a different qualitative behaviour. Two examples of such an order parameter that 
is applicable to traffic flows, can be found in Schadschneider et al. who 
considered nearest neighbour correlations \cite{SCHADSCHNEIDER:02}, and in Jost 
and Nagel who devised a measure of inhomogeneity \cite{JOST:03} (we refer the 
reader to our work in \cite{MAERIVOET:04i} for an example in which they are used 
and compared when tracking phase transitions).

There exists a difference in which a phase transition can express itself. This 
difference is designated by the order of the transition; generally speaking, the 
two most common phase transitions are \emph{first-order} and \emph{second-order 
transitions}. According to \emph{Ehrenfest's classification}, first-order 
transitions have an abrupt, discontinuous change in the order parameter that 
characterises the transition. In contrast to this, the changeover to the new 
phase occurs smoothly for second-order transitions \cite{YANG:52,LEE:52}. Note 
that higher-order phase transitions also exist, e.g., in superconducting 
materials \cite{CRONSTROM:01}.

With respect to the description of regimes in traffic flows, it is commonly 
agreed that there exists a first-order phase transition when going from the 
capacity-flow to the congested regime. The point at which this transition 
occurs, is the critical density. Studying the phase transitions encountered in 
fluid dynamics, there exists a transition from the \emph{laminar} flow (i.e., a 
fluid flowing in layers, each moving at a different velocity) to the 
\emph{turbulent} flow (i.e., the disturbed random and unorganised state in which 
vortices form). However, the transition here is triggered by an increase in the 
velocity of the fluid, as opposed to the transition in traffic flows where a 
change in the density can lead to a cascading instability. In this respect, the 
analogy for traffic flows holds better when comparing them to gas-liquid 
transitions. Here free-flow traffic corresponds to a gaseous phase, in which 
particles are evenly spread out in the system. At the point of the phase 
transition, liquid droplets will form, coagulating together into bigger 
droplets. This leads to a state where both gaseous and liquid phases coexist, 
typically in the form of a big liquid droplet surrounded by gas particles. For 
even higher densities, particles are so close to each other, and the only 
remaining state is the liquid phase 
\cite{EISENBLATTER:98,KRAUSS:99,KAYATZ:01,JOST:02,JOST:03,NAGEL:03}.

In conclusion, we refer the reader to the work of Tamp\`ere, where an excellent 
overview is given, detailing the different traffic flow regimes, their 
transitions, and mechanisms with respect to jamming behaviour \cite{TAMPERE:04}.

		\subsection{Correlations between traffic flow characteristics}

Whereas the previous sections all treated the macroscopic traffic flow 
characteristics on an individual basis, this section considers some of the 
relations between them. We start our discussion with a look at the historic 
origin of fundamental diagrams, after which we shed some light on the different 
classical approaches. The section concludes with some considerations with 
respect to empirical measurements.

			\subsubsection{The historic origin of the fundamental diagram}
			\label{sec:TFT:HistorigOriginOfTheFD}

As in many scientific disciplines, the resulting statements and theories are 
often preceded by an investigation of obtained experimental data, which serves 
as empirical evidence for them. In this line of reasoning, Greenshields was 
among the first to provide --- as far back as 1935 --- a basis for most of the 
classic work on, what are called, \emph{empirical fundamental diagrams}. In his 
seminal paper, he sketched a \emph{linear relation between the density and the 
mean speed}, based on empirically obtained data \cite{GREENSHIELDS:35}:

\begin{equation}
\label{eq:TFT:GreenshieldsKVRelation}
	\overline v = \overline v_{\text{ff}} \left( \one - \frac{k}{k_{j}} \right).
\end{equation}

As can be seen from Greenshields' relation, when increasing the density from 
zero to the jam density $k_{j}$, the mean speed will monotonically decrease from 
the free-flow speed $\overline v_{\text{ff}}$ to zero (note that we dropped the 
`s' or `t' subscript from the mean speed, as it is not sure whether or not 
Greenshields used space- or time-mean speed, respectively). The relation can be 
understood intuitively, by assuming that drivers will tend to slow down in 
crowded traffic, because this naturally gives them more time to react to changes 
(e.g., sudden braking of the lead vehicle). As it is reasonable to assume that 
the mean speed remains unaffected for very low densities, Greenshields 
furthermore flattened the upper-left part of the regression line (corresponding 
to the free-flow speed), although this effect is not incorporated in equation 
\eqref{eq:TFT:GreenshieldsKVRelation}.\\

\begin{figure}[!htb]
	\centering
	\includegraphics[width=\figurewidth]{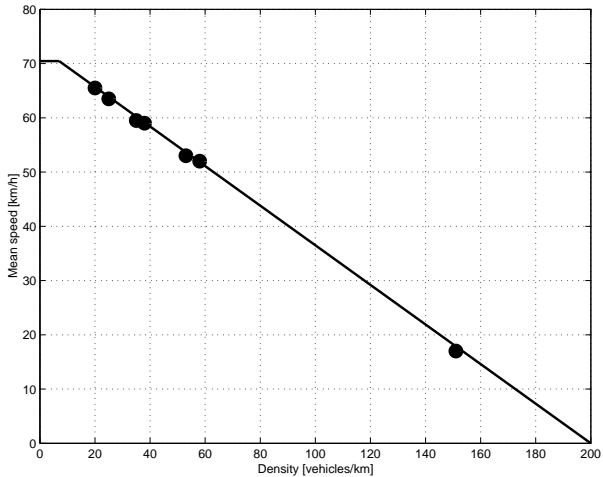}
	\caption{
		Greenshields' original linear relation between the density and the mean 
		speed. Note that the regression line is based on only seven measurements 
		points, and that artificial flattening of its upper-left part (figure based 
		on \cite{LIGHTHILL:55} and \cite{GARTNER:97}).
	}
	\label{fig:TFT:GreenshieldsKVRelation}
\end{figure}

Although Greenshields' derivation of the linear relation between density and 
space-mean speed appears elegant and simple, it should nevertheless be taken 
with a grain of salt. The fact of the matter is that his hypothesis is, as can 
be seen in \figref{fig:TFT:GreenshieldsKVRelation}, based on \emph{only seven 
measurement points}, which comprise aerial observations taken on September 3rd 
(Monday, Labor Day), 1934 \cite{MAY:90}. One of the problems is that these 
observations are \emph{not independent}. An even more serious problem is that 
six of these observations were obtained for free-flow conditions, whereas the 
one single point that indicates congested conditions, was obtained at an 
entirely \emph{different road}, on a \emph{different day} \cite{GARTNER:97}~!\\

Some twenty years later, Lighthill and Whitham developed a theory that describes 
the traffic flows on long crowded roads using a first-order fluid-dynamic model 
\cite{LIGHTHILL:55}. As one of the main ingredients in their theory, they 
postulated the following fundamental hypothesis: \emph{``at any point of the 
road, the flow $q$ is a function of the density $k$''}. They called this 
function the \emph{flow-concentration curve} (recall from section 
\ref{sec:TFT:Occupancy} that density in the past got sometimes referred to as 
concentration).

Continuing their reasoning, Lighthill and Whitham then referred to Greenshields' 
earlier work, relating the space-mean speed to the density, and, by means of 
equation \eqref{eq:TFT:FundamentalRelation}, thus relating the flow to the 
density. The \emph{existence} of the concept of the flow-concentration curve 
mentioned above, was justified on the grounds that it describes traffic 
operating under steady-state conditions, i.e., homogeneous and stationary 
traffic as explained in section \ref{sec:TFT:FundamentalRelation}. In this 
context, the flow-concentration curve therefore describes the \emph{average 
characteristics} of a traffic flow. So Greenshields first fitted a regression 
line to scarce data, after which his functional form seemed to be taken for 
granted for the following seventy years. The key aspect in Lighthill and 
Whitham's (and also Richards' \cite{RICHARDS:56}) approach, lay in the fact that 
they broadened the flow-concentration curve's validity, including also 
conditions of non-stationary traffic. They also stated that, because of e.g., 
changes in the traffic composition, the curve can vary from day to day, or even 
within a day (e.g., rush hours, \ldots). The same statement holds also true when 
considering the flow-concentration curves of different vehicle classes (e.g., 
cars and trucks).

The term \emph{fundamental diagram} itself, is historically based on Lighthill 
and Whitham's \emph{fundamental} hypothesis of the existence of such a 
\emph{one-dimensional} flow-concentration curve. As traffic engineers grew 
accustomed to the graphical representation of this curve, they started talking 
about the \emph{diagram} that represents it, i.e., the 'fundamental diagram' 
\cite{HAIGHT:63}.\\

\sidebar{
	In its original form, the fundamental diagram represents an \emph{equilibrium 
	relation} between flow and density, denoted by $q_{e}(k)$. But note that, 
	because of the fundamental relation of traffic flow theory (see section 
	\ref{sec:TFT:FundamentalRelation}), is it equally justified to talk about the 
	$\overline v_{s_{e}}(k)$ or the $\overline v_{s_{e}}(q)$ fundamental diagrams. 
	Due to this equilibrium property, the traffic states (i.e., the density, flow, 
	and space-mean speed) can be thought of as `moving' over the fundamental 
	diagrams' curves.
}\\

			\subsubsection{The general shape of a fundamental diagram}
			\label{sec:TFT:GeneralShapeOfAFD}

We now give an overview of some of the qualitative features of the different 
possible fundamental diagrams, representing the equilibrium relations between 
density, space-mean speed, and average space headway, and flow. Note that in 
each example, we consider a \emph{possible} fundamental diagram, as they can 
take on many (functional) shapes.

\textbf{Space-mean speed versus density}\\
We start our discussion based on the equilibrium relation between space-mean 
speed and density, i.e., the $\overline v_{s_{e}}(k)$ fundamental diagram. The 
main reason for starting here, is the fact that this diagram is the easiest to 
understand intuitively. Complementary to the example of Greenshields in 
\figref{fig:TFT:GreenshieldsKVRelation}, we give a small overview of its most 
prominent features:

\begin{itemize}
	\item the density is restricted between 0 and the maximum density, i.e., 
	the jam density $k_{j}$,
	\item the space-mean speed is restricted between 0 and the maximum 
	average speed, i.e., the free-flow speed $\overline v_{\text{ff}}$,
	\item as density increases, the space-mean speed monotonically 
	decreases,
	\item there exists a small range of low densities, in which the space-mean
	speed remains unaffected and corresponds more or less to the free-flow speed,
	\item and finally, the flow (equal to density times space-mean speed), 
	can be derived as the area demarcated by a rectangle who's lower-left 
	and upper-right corners are the origin and a point on the fundamental 
	diagram, respectively.
\end{itemize}

\begin{figure}[!htb]
	\centering
	\psfrag{0}[][]{\psfragstyle{\zero}}
	\psfrag{vs}[][]{\psfragstyle{$\overline v_{s}$}}
	\psfrag{vff}[][]{\psfragstyle{$\overline v_{\text{ff}}$}}
	\psfrag{k-1}[][]{\psfragstyle{$k^{-\one} \propto \overline h_{s}$}}
	\psfrag{kj-1}[][]{\psfragstyle{$k_{j}^{-\one}$}}
	\psfrag{kc-1}[][]{\psfragstyle{$k_{c}^{-\one}$}}
	\includegraphics[width=\figurewidth]{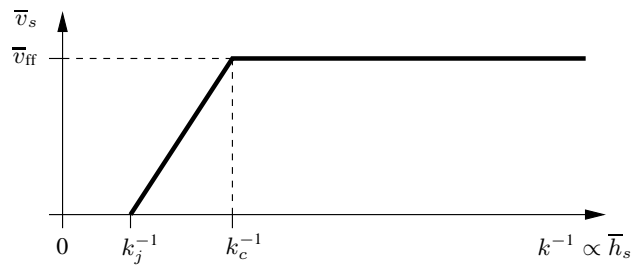}
	\caption{
		A fundamental diagram relating the average space headway $\overline h_{s}$ 
		to the space-mean speed $\overline v_{s}$. Note that the average space 
		headway is proportional to the inverse of the density, i.e., $k^{-\one}$.
	}
	\label{fig:TFT:AverageSpaceHeadwaySMSFD}
\end{figure}

\textbf{Space-mean speed versus average space headway}\\
Microscopic and macroscopic traffic flow characteristics are related to each 
other by means of equations \eqref{eq:TFT:DensitySpaceHeadwayRelation} and 
\eqref{eq:TFT:FlowTimeHeadwayRelation}. According to the former, density $k$ is 
inversely proportional to the average space headway $\overline h_{s}$. We can 
therefore derive a fundamental diagram, similar to the previous one, by 
substituting the density with the average space headway. As as result, the 
abscissa gets `inverted', resulting in the fundamental diagram as shown in 
\figref{fig:TFT:AverageSpaceHeadwaySMSFD}.

The interesting features of this type of fundamental diagram, can be 
summed as follows:

\begin{itemize}
	\item the curve starts not in the origin, but at $k_{j}^{-\one}$, 
	corresponding to the average space headway when the jam density is 
	reached (i.e., all vehicles are standing nearly bumper to bumper),
	\item as the average space headway increases, its inverse (the density) 
	decreases, and the space-mean speed increases,
	\item the space-mean speed continues to rise with an increasing average 
	space headway, until it reaches the maximum average speed, i.e., the 
	free-flow speed $\overline v_{\text{ff}}$; this happens at the inverse 
	of the critical density $k_{c}^{-\one}$,
	\item from then on, the space-mean speed remains constant with 
	increasing average space headway.
\end{itemize}

The above features can be understood intuitively: at large average space 
headways, a driver experiences no influence from its direct frontal leader. 
However, there exists a point at which the driver comes `close enough' to this 
leader (i.e., in crowded traffic), so that its speed will decrease. This slowing 
down will continue to persist as traffic gets more dense (this the same 
reasoning behind Greenshields' derivation in section 
\ref{sec:TFT:HistorigOriginOfTheFD}).\\

\begin{figure}[!htb]
	\centering
	\psfrag{0}[][]{\psfragstyle{\zero}}
	\psfrag{q}[][]{\psfragstyle{$q$}}
	\psfrag{qcap}[][]{\psfragstyle{$q_{\text{cap}}$}}
	\psfrag{k}[][]{\psfragstyle{$k$}}
	\psfrag{kj}[][]{\psfragstyle{$k_{j}$}}
	\psfrag{kc}[][]{\psfragstyle{$k_{c}$}}
	\psfrag{vs}[][]{\psfragstyle{$\overline v_{s}$}}
	\psfrag{w}[][]{\psfragstyle{$w$}}
	\psfrag{free-flow}[][]{\psfragstyle{\emph{free-flow}}}
	\psfrag{congested}[][]{\psfragstyle{\emph{congested}}}
	\includegraphics[width=\figurewidth]{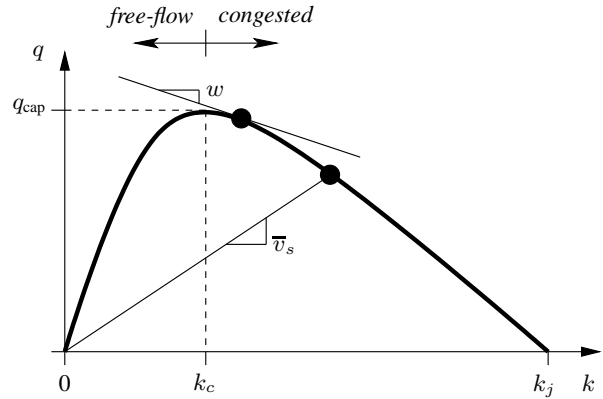}
	\caption{
		A fundamental diagram relating the density $k$ to the flow $q$. The capacity 
		flow $q_{\text{cap}}$ is reached at the critical density $k_{c}$. The 
		space-mean speed $\overline v_{s}$ for any point on the curve, is defined as 
		the slope of the line through that point and the origin. Taking the slope of 
		the tangent to points on the curve, gives the characteristic wave speed $w$.
	}
	\label{fig:TFT:DensityFlowFD}
\end{figure}

\textbf{Flow versus density}\\
Probably the most encountered form of a fundamental diagram, is that of flow 
versus density. Its origins date back to the seminal work of Lighthill and 
Whitham who, as described earlier, referred to it as the flow-concentration 
curve. An example of the $q_{e}(k)$ fundamental diagram is depicted in 
\figref{fig:TFT:DensityFlowFD}.

Noteworthy features of this type of fundamental diagram are:

\begin{itemize}
	\item for moderately low densities (i.e., below the critical density 
	$k_{c}$), the flow increases more or less linearly (this is called the 
	\emph{free-flow branch} of the fundamental diagram),
	\item near the critical density $k_{c}$, the fundamental diagram can bend 
	slightly, due to faster vehicles being obstructed by slower vehicles, 
	thereby lowering the free-flow speed \cite{NEWELL:93b},
	\item at the critical density $k_{c}$, the flow reaches a maximum, called 
	the capacity flow\footnote{Note that this capacity flow is not an extreme 
	value, i.e., it can be different from the maximum observed flow. The 
	reason is that, with respect to the nature of the fundamental diagram, the 
	capacity flow is taken to be an \emph{average} value 
	\cite{GREENSHIELDS:35,LIGHTHILL:55}.} $q_{\text{cap}}$,
	\item in the congested regime (i.e., for densities higher than the 
	critical density), the flow starts to degrade with increasing density, 
	until the jam density $k_{j}$ is reached and traffic comes to a stand 
	still, resulting in a zero flow (this is called the \emph{congested 
	branch} of the fundamental diagram),
	\item the space-mean speed $\overline v_{s}$ for any point on the 
	$q_{e}(k)$ fundamental diagram, can be found as the slope of the line 
	through that point and the origin,
\end{itemize}

There is one more piece of information revealed by the $q_{e}(k)$ fundamental 
diagram: When taking the slope of the tangent in any point of the diagram, we 
obtain what is called the \emph{kinematic wave speed}. These speeds $w$ 
correspond to \emph{shock waves} encountered in traffic flows (e.g., the 
stop-and-go waves). As can be seen from the figure, the shock waves travel 
forwards, i.e., \emph{downstream}, in free-flow traffic ($w \geq \zero$), but 
backwards, i.e., \emph{upstream}, in congested traffic ($w \leq \zero$).

The above shape of the $q_{e}(k)$ fundamental diagram is just one possibility. 
There exist many different flavours, originally derived by traffic engineers 
seeking a better fit of these curves to empirical data. After the work of 
Greenshields, another functional form --- based on a logarithm --- was proposed 
by Greenberg \cite{GREENBERG:59}. Another possible form was introduced by 
Underwood \cite{UNDERWOOD:61}. All of the previous diagrams are called 
\emph{single-regime models}, because they formulate only one relation between 
the macroscopic traffic flow characteristics for the entire range of densities 
(i.e., traffic flow regimes) \cite{MAY:90}. In contrast to this, Edie started 
developing \emph{multi-regime models}, allowing for discontinuities and a better 
fit to empirical data coming from different traffic flow regimes \cite{EDIE:61}. 
We refer the reader to the work of Drake et al. \cite{DRAKE:67} and the book of 
May \cite{MAY:90} for an extensive comparison and overview of these different 
modelling approaches (note that Drake et al. used time-mean speed).

During the last two decades, other, sometimes more sophisticated, functional 
relationships between density and flow have been proposed. Examples are the work 
of Smulders who created a non-differentiable point at the critical density in a 
two-regime fundamental diagram \cite{SMULDERS:89}, the METANET model of Messmer 
and Papageorgiou who's single-regime fundamental diagram contains an inflection 
point near the jam density \cite{MESSMER:90}, the work of De Romph who 
generalised Smulders' functional description of his two-regime fundamental 
diagram \cite{DEROMPH:94}, the typical triangular shape of the fundamental 
diagram introduced by Newell, resulting in only two possible values for the 
kinematic wave speed $w$ \cite{NEWELL:93b}, \ldots As can be seen, these 
fundamental diagrams sometimes take on non-concave forms, depending on the 
existence of inflection points in the functional relation between flow and 
density. In general, they can be convex, concave, (dis)continuous, 
piecewise-linear, everywhere differentiable, have inflection points, \ldots 
Variations in shape will continu to be proposed, as it is for certain that there 
is no general consensus among traffic engineers regarding the correct shape of 
this fundamental diagram. To illustrate this, a more exotic approach is based on 
\emph{catastrophe theory}, which is, in a sense, a three-dimensional model that 
jointly treats density, flow, and space-mean speed. Acha-Daza and Hall applied 
the technique, resulting in a satisfactory fit with empirical data 
\cite{ACHADAZA:94}.

The most extreme argument with respect to the shape of the fundamental diagram, 
came from Kerner who questioned its validity, and consequently rejected it 
altogether by replacing it with his fundamental hypothesis of \emph{three-phase 
traffic flow theory} (refer to section \ref{sec:TFT:Kerners3PT} for more 
details) \cite{KERNER:04}.\\

\begin{figure}[!htb]
	\centering
	\psfrag{0}[][]{\psfragstyle{\zero}}
	\psfrag{q}[][]{\psfragstyle{$q$}}
	\psfrag{qcap}[][]{\psfragstyle{$q_{\text{cap}}$}}
	\psfrag{vs}[][]{\psfragstyle{$\overline v_{s}$}}
	\psfrag{vff}[][]{\psfragstyle{$\overline v_{\text{ff}}$}}
	\includegraphics[width=\figurewidth]{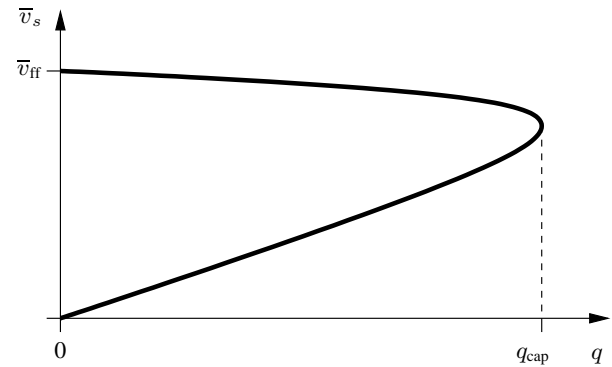}
	\caption{
		A fundamental diagram relating the flow $q$ to the space-mean speed 
		$\overline v_{s}$. The capacity flow $q_{\text{cap}}$ is located at the 
		right edge of the diagram, i.e., it is defined as the maximum average flow. 
		Note that there are two possible speeds associated with each value of the 
		flow.
	}
	\label{fig:TFT:FlowSMSFD}
\end{figure}

\textbf{Space-mean speed versus flow}\\
An often spotted shape is that of the $\overline v_{s_{e}}(q)$ fundamental 
diagram, depicted in \figref{fig:TFT:FlowSMSFD}. As opposed to the earlier 
discussed $\overline v_{s_{e}}(k)$ fundamental diagram, the space-mean speed 
versus flow curve no longer embodies a function in the strict mathematical 
sense: for each value of the flow, there exists two different mean speeds, 
namely one in the free-flow regime (upper branch) and one in the congested 
regime (lower branch).

Some people, e.g., economists who use the flow to represent traffic demand, find 
this kind of fundamental diagram easy to cope with. But in our opinion, we are 
convinced however, that this diagram is rather difficult to understand at first 
sight. We believe the $\overline v_{s_{e}}(k)$ fundamental diagram is a much 
better candidate, because density can intuitively be understood as a measure for 
how crowded traffic is, as opposed to some flow giving rise to two different 
values for the space-mean speed.

As a final comment, we would like to point out that the previously discussed 
bivariate functional relationships between the traffic flow characteristics 
(e.g., density and flow), are based on observations. More importantly, this 
means that there is \emph{no direct causal relation} assumed between any two 
variables. Fundamental diagrams sketch \emph{only possible correlations}, 
implying that the \emph{nature of the transitions} between different traffic 
regimes thus remains to be explored (see section \ref{sec:TFT:PhaseTransitions} 
for a discussion).

			\subsubsection{Empirical measurements}

As mentioned earlier, the fundamental diagrams discussed in the previous section 
represent equilibrium relations between the macroscopic traffic flow 
characteristics of section \ref{sec:TFT:MacroCharacteristics}. In sharp contrast 
to this, real empirical measurements from detector stations do not describe such 
nice one-dimensional curves corresponding to the functional relationships.

\begin{figure}[!htb]
	\centering
	\includegraphics[width=\figurewidth]{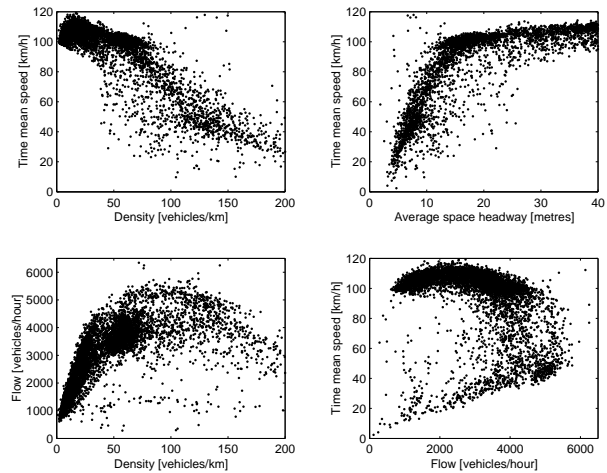}
	\caption{
		Illustrative scatter plots of the relations between traffic flow 
		characteristics as measured by video camera CLO3 located at the E17 
		three-lane motorway near Linkeroever, Belgium. The measured occupancies were 
		converted into densities, the time-mean speed remained unchanged. Shown are 
		scatter plots of a ($k$,$\overline v_{t}$) diagram (\emph{top-left}), a 
		($\overline h_{s}$,$\overline v_{t}$) diagram (\emph{top-right}), a 
		($k$,$q$) diagram (\emph{bottom-left}), and a ($q$,$\overline v_{t}$) 
		diagram (\emph{bottom-right}).
	}
	\label{fig:TFT:CLO3FDs}
\end{figure}

As an illustrative example, we provide some scatter plots in 
\figref{fig:TFT:CLO3FDs}. The shown data comprises detector measurements (the 
sampling interval was one minute) during the entire year 2003; they were 
obtained by means of a video camera \cite{VVC:03} located at the E17 three-lane 
motorway near Linkeroever\footnote{The detector station is called CLO3, which is 
an acronym for `Camera Linkeroever'.}, Belgium. Because of the nature of this 
data, we only obtained flows, occupancies, and time-mean speeds. After 
calculating the average vehicle length, the occupancies were converted into 
densities using equation \eqref{eq:TFT:OccupancyDensity}. Using these recorded 
time series, we then constructed scatter plots of the density, time-mean speed, 
flow, and average space headway. Note that no substantial changes are introduced 
in these plots due to e.g., our using of densities calculated from occupancies, 
instead of using real measured densities.

As the dimension of time is removed in these scatter plots, Daganzo calls them 
\emph{time-independent models} \cite{DAGANZO:97}. It is important to understand 
that \emph{these scatter plots are not fundamental diagrams}, because the latter 
represent one-dimensional equilibrium curves. According to Helbing, a better 
designation would be \emph{regression models} \cite{HELBING:97e}. In this 
dissertation, we introduce a terminology based on \emph{phase spaces} (or 
equivalently \emph{state spaces}), resulting in e.g., the ($k$,$q$) diagram 
(note that we dropped the adjective `fundamental').

In reality, traffic is not homogeneous, nor is it stationary, thus having the 
effect of a large amount of scatter in the presented diagrams. In free-flow 
traffic, interactions between vehicles are rare, and their small local 
disturbances have no significant effects on the traffic stream. As a result, all 
points are somewhat densely concentrated along a line --- representing the 
free-flow speed --- in all four diagrams. However, in the congested regime, a 
wide range of scatter is visible due to the interactions between vehicles. 
Furthermore, vehicle accelerations and decelerations lead to large fluctuations 
in the traffic stream, as can be seen by the thin, but large, cloud of data 
points. The effect is especially pronounced for intermediate densities, leading 
to large fluctuations in the time-mean speed and flow.\\

\sidebar{
	The occurrence of all this scatter in the data, leads some traffic engineers 
	to question the validity of the fundamental diagram. More specifically, the 
	behaviour in congested traffic seems ill-defined to some. As stated earlier, 
	Kerner is the most intense opponent in this debate, as he outright rejects 
	Lighthill and Whitham's hypothesis that remained popular over the last fifty 
	years. Despite this criticism, the fundamental diagram remains, to the 
	majority of the community, a fairly accurate description of the average 
	behaviour of a traffic stream. Cassidy even provided quantifiable evidence of 
	the existence of well-defined bivariate relations between traffic flow 
	characteristics. The key here was to separate stationary periods from 
	non-stationary ones in the detector data (i.e., stratifying it) 
	\cite{DAGANZO:97,CASSIDY:98}. Prior work of Del Castillo and Ben\'itez 
	resulted in a more mathematically justified method, for fitting empirical 
	curves in data regions of stationary traffic, after construction of a rigid 
	set of properties that all fundamental diagrams should satisfy 
	\cite{DELCASTILLO:95,DELCASTILLO:95b}.
}\\

As a final note, we remark that the distribution of the cloud-like data points 
of the diagrams in \figref{fig:TFT:CLO3FDs}, is a result of various kinds of 
phenomena. First and foremost, there is the heterogeneity in the traffic 
composition (fast passenger cars, slow trailer trucks, \ldots). Secondly, as 
already mentioned, the non-stationary behaviour of traffic introduces a 
significant amount of scatter in the congested regime. Thirdly, each scatter 
plot is dependent on the type of road, and the time of day at which the 
measurements were collected. In this respect, the influence of (changing) 
weather conditions is not to be underestimated (e.g., rain fall results in 
different diagrams). In conclusion, it is clear that if we want these scatter 
plots to better fit the fundamental diagrams, all data points should be 
collected under similar conditions. Even more so, the relative location on the 
road at which the data points were recorded plays a significant role: e.g., a 
jam that propagates upstream, passing an on-ramp will show different effects, 
depending on where the observations were gathered (upstream, right at, or 
downstream of the on-ramp) and on whether or not the particular bottleneck was 
active \cite{MAY:90}.

		\subsection{Capacity drop and the hysteresis phenomenon}
		\label{sec:TFT:CapacityDropAndHysteresis}

In the early sixties, traffic engineers frequently observed a discontinuity in 
the measurements near the capacity flow. To this end, Edie proposed a two-regime 
model that included such a discontinuity at the critical density \cite{EDIE:61}. 
Nowadays, this typical form of the $q_{e}(k)$ fundamental diagram is known as a 
\emph{reversed lambda shape} (the name was originally suggested by Koshi et al. 
\cite{KOSHI:83}).

\begin{figure}[!htb]
	\centering
	\psfrag{q}[][]{\psfragstyle{$q$}}
	\psfrag{qcap}[][]{\psfragstyle{$q_{\text{cap}}$}}
	\psfrag{qout}[][]{\psfragstyle{$q_{\text{out}}$}}
	\psfrag{k}[][]{\psfragstyle{$k$}}
	\psfrag{kout}[][]{\psfragstyle{$k_{\text{out}}$}}
	\psfrag{kc}[][]{\psfragstyle{$k_{c}$}}
	\psfrag{kj}[][]{\psfragstyle{$k_{j}$}}
	\psfrag{0}[][]{\psfragstyle{\zero}}
	\psfrag{(1)}[][]{\psfragstyle{(\one)}}
	\psfrag{(2)}[][]{\psfragstyle{(\two)}}
	\psfrag{(3)}[][]{\psfragstyle{(\three)}}
	\includegraphics[width=\halffigurewidth]{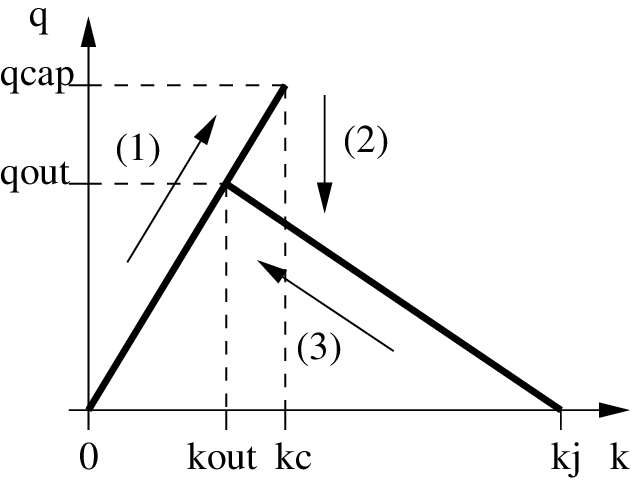}
	\hspace*{\figureseparation}
	\psfrag{0}[][]{}
	\includegraphics[width=\halffigurewidth]{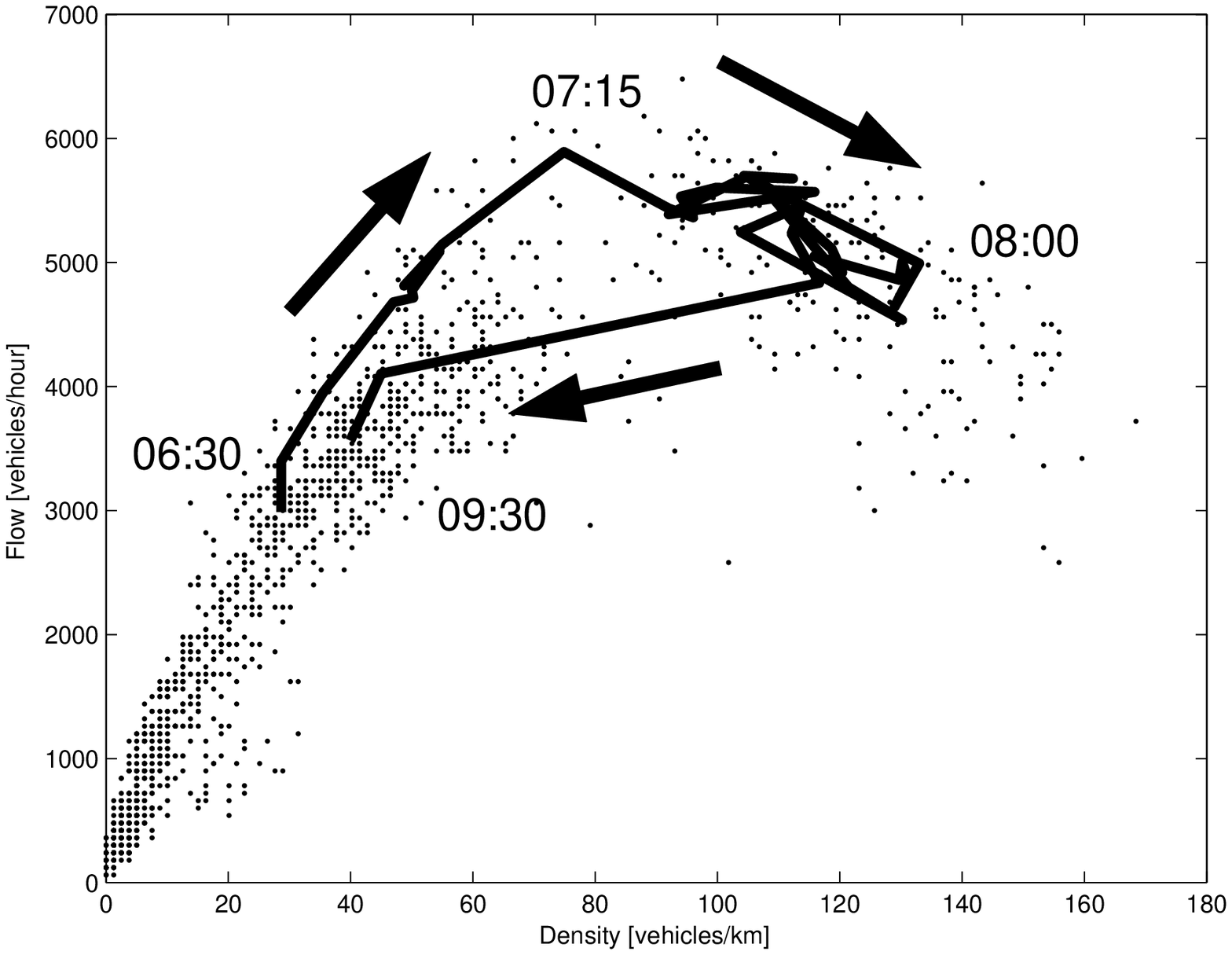}
	\caption{
		\emph{Left:} the typical inverted $\lambda$ shape of the ($k$,$q$) 
		fundamental diagram, showing a capacity drop from $q_{\text{cap}}$ to below 
		$q_{\text{out}} \ll q_{\text{cap}}$ (i.e., the queue discharge flow). The 
		hysteresis effect occurs when going from the congested to the free-flow 
		branch, as indicated by the three arrows (1) -- (3). \emph{Right:} a 
		($k$,$q$) diagram based on empirical data of one day, obtained by video 
		camera CLO3, at the E17 three-lane motorway near Linkeroever, Belgium. The 
		black dots denote minute measurements, whereas the thick solid line 
		represents the time-traced evolution of traffic conditions. The observed 
		hysteresis loop was based on consecutive 5-minute intervals covering a 
		period that encompasses the morning rush hour between 06:30 and 09:30.
	}
	\label{fig:TFT:HysteresisCLO3}
\end{figure}

An example of such a reversed $\lambda$ fundamental diagram, is shown in the 
left part of \figref{fig:TFT:HysteresisCLO3}. Note however, that the depicted 
discontinuity apparently leads to \emph{overlapping} branches of the free-flow 
and congested regimes, resulting in a \emph{multi-valued} fundamental diagram.

Considering the left part of \figref{fig:TFT:HysteresisCLO3}, it appears the 
flow can take on two different values (hence the name `two-regime, two-capacity' 
model) depending on the traffic conditions, i.e., whether traffic is moving from 
the free-flow to the congested regime on the equilibrium curve or vice versa. In 
order to comprehensively understand this \emph{hysteretic} behaviour, we 
consider the following intuitive sequence of events:

\begin{quote}
	\textbf{(1)} In the free-flow regime, the flow steadily rises with increasing 
	density, small perturbations in the traffic flow have no significant effects 
	(see section \ref{sec:TFT:FreeFlowTraffic}).

	\textbf{(2)} At the critical density $k_{c}$, traffic is said to be 
	\emph{metastable}: for small disturbances, traffic is stable, but when these 
	disturbances are sufficiently large, they can lead to a cascading effect (see 
	section \ref{sec:TFT:CapacityFlowTraffic}), resulting in a breakdown of 
	traffic and kicking it onto the congested branch. The state of capacity flow 
	at $q_{\text{cap}}$ is destroyed, due to a sudden decrease of the flow, called 
	the \emph{capacity drop}.

	\textbf{(3)} In order to recover from the congested to the free-flow regime, the 
	traffic density has to be \emph{reduced substantially} (in comparison with the 
	reverse transition), i.e., well below the critical density $k_{c}$. After this 
	recovery, the flow will \emph{not} be equal to $q_{\text{cap}}$, but to 
	$q_{\text{out}} \ll q_{\text{cap}}$, which is called the \emph{outflow from a 
	jam} or the \emph{queue discharge capacity}.
\end{quote}

The above sequence signifies a \emph{hysteresis loop} in the flow versus density 
fundamental diagram: going from the free-flow to the congested regime occurs via 
the capacity flow, but the reverse transition proceeds via another way. The 
phenomenon was first observed by Treiterer and Meyers, who used aerial 
photography to calculate densities and space-mean speeds, extracted from a 
platoon of moving vehicles \cite{TREITERER:74}. Hall et al. later observed a 
similar phenomenon \cite{HALL:86}.

The right part of \figref{fig:TFT:HysteresisCLO3} shows a ($k$,$q$) diagram, 
obtained with empirical data collected at Monday September 10th, 2001. The data 
was recorded by video camera CLO3, at the E17 three-lane motorway near 
Linkeroever, Belgium. The small dots represent minute-based measurements, 
whereas the thick solid line represents the time-traced evolution of traffic 
conditions. The observed hysteresis loop was based on consecutive 5-minute 
intervals covering a period that encompasses the morning rush hour between 06:30 
and 09:30.

Zhang is among the few who try to give a possible \emph{rigourous mathematical 
explanation} for the occurrence of this hysteresis phenomenon \cite{ZHANG:99b}. 
His exposition is based on the behaviour of individual drivers during 
car-following: central to his interpretation is the existence of an 
\emph{asymmetry} between accelerating and decelerating vehicles (a related 
notion was already explored by Newell back in 1963 \cite{NEWELL:63}). The former 
are associated with larger space headways, whereas the latter typically have 
smaller space headways. Both observations can be understood when considering the 
characteristic `harmonica' effect of a string of consecutive vehicles: when the 
next stop-and-go wave is encountered, a driver is more alert as he typically has 
to brake rather hard in order to avoid a collision. But once this wave has 
passed, a driver gets more relaxed, resulting in a larger response time when 
applying the gas pedal. The deceleration reaction leads to a sudden decrease of 
the space headway, whereas the acceleration reaction leads to a gradually 
developing larger space headway. To this end, Zhang introduces three distinct 
traffic phases, respectively called the \emph{acceleration phase}, the 
\emph{deceleration phase}, and a \emph{strong equilibrium} (indicating a 
constant speed). Because the space headway is thus treated differently under 
these qualitatively different circumstances, the result is that there are now 
different functional relations for the $\overline v_{s_{e}}(\overline h_{s})$ 
fundamental diagram. As a consequence, a hysteresis loop can appear in the 
(density,flow) state space. Note that Zhang's work describes a \emph{continuous} 
loop in state space, whereas in most cases hysteresis is assumed to follow a 
discontinuous fundamental diagram. Furthermore, as there are three different 
ways for vehicles to reside in a traffic stream (i.e., Zhang's traffic phases), 
there are now three different capacities related to these conditions; it is the 
capacity under a stationary equilibrium flow that should be considered as the 
ideal capacity of a roadway \cite{ZHANG:01b}.

Note that depending on the location where the traffic stream measurements were 
performed, the transition from the free-flow to the congested regime and vice 
versa does not always have to pass via the capacity flow. Instead, observations 
can indicate that the traffic state can jump abruptly from one branch to another 
in the diagram \cite{GARTNER:97}. A possible explanation is that upstream of a 
jam, vehicles arrive with high speeds, resulting in strong decelerations; a 
detector station located at this point would observe traffic jumping from the 
uncongested branch immediately to the congested branch, without necessarily 
having to pass via the capacity \cite{NEWELL:82}. This has led Hall et al. to 
believe the reversed lambda shape is more correctly replaced by a continuous but 
non-differentiable \emph{inverted V shape} \cite{HALL:86}.

Continuing this latter train of thought, Daganzo believes that many of these 
`extravagant' phenomena (e.g., a multi-valued fundamental diagram) are uncalled 
for. Applying the stratification methodology of Cassidy \cite{CASSIDY:98}, the 
scatter in the empirical data may vanish, restoring a smooth continuous 
equilibrium relation between density and flow. One way of explaining the high 
tip of the lambda, is to assume that it is caused by statistical fluctuations 
that comprise platoons of densely packed vehicles \cite{DAGANZO:97}.\\

\sidebar{
	During the last seventy years, there has been a continuing quest to find the 
	`correct' form of the fundamental diagrams. In this respect, we like to stress 
	the fact that `only looking at the measurements' is not sufficient: traffic 
	engineers wanting to mine the gigabytes of empirical data, should always look 
	at the global picture. This means that the typical driving patterns, \emph{as 
	well as the local geometry/infrastructure}, should also be taken into account, 
	so that the local measurements can be interpreted with respect to the traffic 
	flow dynamics. If this is neglected, the danger exists that traffic is only 
	sampled at discrete locations, giving a sort of `truncated' view of the 
	occurring dynamical processes.\\

	Finally, we like to agree with Zhang's comments: the root cause of most of the 
	differences in the construction of fundamental diagrams, is the erroneous 
	treatment of data (e.g., mixing data stemming from different traffic flow 
	regimes) \cite{ZHANG:99b}. Because fundamental diagrams imply the notion of an 
	equilibrium, care should be taken when using the data, i.e., only considering 
	stationary periods after removing the transients.
}\\

		\subsection{Kerner's three-phase theory}
		\label{sec:TFT:Kerners3PT}

In the mid-nineties, Kerner and other fellow researchers, studied various 
traffic flow measurements stemming from detector stations along German 
motorways. Initially, they agreed with the classic notion of Lighthill and 
Whitham's fundamental hypothesis of the existence of one-dimensional equilibrium 
relation between the macroscopic traffic flow characteristics (see section 
\ref{sec:TFT:HistorigOriginOfTheFD} for more details). However, upon discovery 
of a rich and complex set of empirical tempo-spatial patterns in congested 
traffic flow, Kerner decided to abolish this hypothesis, as it could not 
adequately capture all of these observed patterns. As a consequence, Kerner 
rejects \emph{all} traffic flow theories and models that are based on this 
one-dimensional equilibrium relation \cite{KERNER:04}.

In the search for a more correct theory that could accurately describe empirical 
traffic flow observations, Kerner developed what is known as the 
\emph{three-phase theory} of traffic flow.

			\subsubsection{Free flow, synchronised flow, and wide-moving jam}
			\label{sec:TFT:Kerners3PhasesExplained}

In section \ref{sec:TFT:TrafficFlowRegimes}, we elaborated on a classic approach 
to traffic flow, general assuming two qualitatively different regimes, namely 
free-flow and congested traffic. Based on empirical findings, Kerner and Rehborn 
in 1996 proposed three different regimes, separating the congested regime into 
two other regimes. This led them to the introduction of the following regimes 
\cite{KERNER:96b}:

\begin{itemize}
	\item free flow,
	\item synchronised flow,
	\item and wide-moving jam.
\end{itemize}

The main difference between \emph{synchronised flow} and the \emph{wide-moving 
jam}, is that in the former low speeds but high flows (comparable to free-flow 
traffic) can be observed, whereas in the latter both low speeds and low flows 
are observed. The description by the term `synchronised' was based on the 
discovery that the time series of flows, densities, and mean speeds exhibited 
large degrees of correlation among neighbouring lanes. And although synchronised 
flow is treated as a form of congestion, it nevertheless is characterised by a 
high continuous flow. Furthermore, a typical tempo-spatial region of 
synchronised flow has a fixed downstream front (that could be located at a 
bottleneck's position), whereas both the upstream and downstream fronts of a 
wide-moving jam can propagate undisturbed in the upstream direction of a traffic 
stream \cite{KERNER:96}.

Kerner distinguishes several congestion patterns with respect to traffic flows. 
A first typical pattern is a \emph{synchronised-flow pattern} (SP), which can be 
further classified as a \emph{moving SP} (MSP), a \emph{widening SP} (WSP), and 
a \emph{localised SP} (LSP). An SP can only contain synchronised flow; as we 
will shortly mention in section \ref{sec:TFT:FSJTransitions}, a moving jam can 
only occur inside such an SP. When such a jam transforms into a wide-moving jam, 
the resulting pattern is called a \emph{general pattern} (GP); a GP therefore 
contains both synchronised flow and wide-moving jams. Just as with the SP, there 
exist different types of GP. These are a \emph{dissolving GP} (DGP), a \emph{GP 
under weak congestion}, and a \emph{GP under strong congestion}. A final often 
encountered pattern occurs when two bottlenecks are spatially close to each 
other, resulting in what is called an \emph{expanded congested pattern} (EP).

Taking the above considerations into account, the discovery and distinction 
between both types of congested traffic patterns should be made on the basis of 
tempo-spatial plots of the speed, rather than the flow (because the flow in 
synchronised traffic is difficult to differentiate from that of free-flow 
traffic) \cite{KERNER:04}. To this end, Kerner et al. developed two applications 
that are capable of accurately estimating, automatically tracking, and reliably 
predicting the above mentioned congested traffic patterns. Their models are the 
\emph{Forecasting of Traffic Objects} (FOTO) and \emph{Automatische 
StauDynamikAnalyse} (ASDA) \cite{KERNER:01}.

			\subsubsection{Fundamental hypothesis of three-phase traffic theory}
			\label{sec:TFT:ThreePhaseTrafficTheoryFundamentalHypothesis}

\begin{figure}[!htb]
	\centering
	\psfrag{0}[][]{\psfragstyle{\zero}}
	\psfrag{q}[][]{\psfragstyle{$q$}}
	\psfrag{qcap}[][]{\psfragstyle{$q_{\text{cap}}$}}
	\psfrag{qout}[][]{\psfragstyle{$q_{\text{out}}$}}
	\psfrag{k}[][]{\psfragstyle{$k$}}
	\psfrag{kj}[][]{\psfragstyle{$k_{j}$}}
	\psfrag{kout}[][]{\psfragstyle{$k_{\text{out}}$}}
	\psfrag{kc}[][]{\psfragstyle{$k_{c}$}}
	\psfrag{F}[][]{\psfragstyle{$F$}}
	\psfrag{S}[][]{\psfragstyle{$S$}}
	\psfrag{J}[][]{\psfragstyle{$J$}}
	\includegraphics[width=\figurewidth]{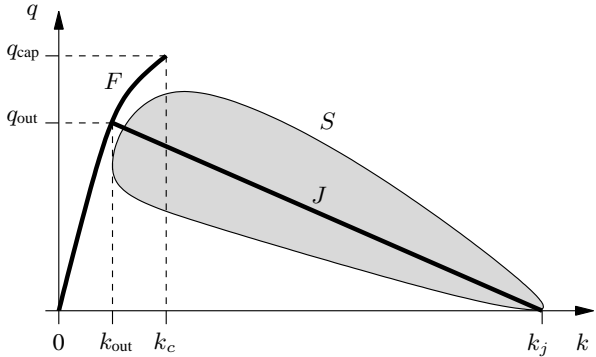}
	\caption{
		The flow versus density relation according to Kerner's three-phase traffic 
		theory. The curve of free flow (denoted by $F$) is reminiscent of 
		observations in the classic free-flow regime. It levels of a bit towards the 
		capacity flow $q_{\text{cap}}$ at the critical density $k_{c}$. As a result 
		of Kerner's fundamental hypothesis, the region of synchronised flow (denoted 
		by $S$) covers a large two-dimensional part of the density-flow phase space. 
		It is intersected by the line $J$, denoting the steady propagation of 
		wide-moving jams. The line $J$ also intersects the curve of free flow in the 
		outflow from a jam $q_{\text{out}} \ll q_{\text{cap}}$ at the associated 
		density $k_{\text{out}}$.
	}
	\label{fig:TFT:KernerDensityFlow}
\end{figure}

Central to Kerner's theory, is the \emph{fundamental hypothesis of three-phase 
traffic theory}, which basically states that hypothetical steady states of 
synchronised flow, cover a \emph{two-dimensional region} in a flow versus 
density diagram (as opposed to the classic notion of a one-dimensional 
equilibrium relation). An example of such a diagram can be seen in 
\figref{fig:TFT:KernerDensityFlow}.

In the flow versus density diagram in \figref{fig:TFT:KernerDensityFlow}, the 
three regimes are depicted: the curve of free flow (denoted by $F$), the region 
of synchronised flow (denoted by $S$) and the wide-moving jam (denoted by the 
empirical line $J$). Just as in the classic fundamental diagrams, the 
observations in free-flow traffic lie on a sharp line that linearly increases 
the flow with higher densities (note the levelling of the curve near the 
capacity flow $q_{\text{cap}}$ associated with the critical density $k_{c}$). 
The region of synchronised flow spans a large part of the density-flow phase 
space; an important remark here is that consecutive measurement points are 
scattered within this region, meaning that an increase in the flow can happen 
with both higher \emph{and} lower densities (as opposed to the free-flow regime) 
\cite{KERNER:96b}.

The characteristic line $J$ denotes the steady, undisturbed propagation of 
wide-moving jams. Its slope corresponds to the speed of a wide-moving jam's 
downstream front, which typically lies around $w \approx$ -15~km/h 
\cite{KERNER:96}. The upper-left point of the line $J$ is located at a density 
$k_{\text{out}}$ corresponding to the outflow $q_{\text{out}} \ll 
q_{\text{cap}}$ from a wide-moving jam. This is an illustration of the capacity 
drop phenomenon, elucidated in section \ref{sec:TFT:CapacityDropAndHysteresis}. 
The line $J$ is defined as follows:

\begin{equation}
	q(k) = \frac{\one}{T} \left ( \one - \frac{k}{k_{\text{jam}}} \right ),
\end{equation}

with $T$ the time gap in congested traffic flows; it is used to tune the outflow 
from a jam. Because wide-moving jams travel undisturbed, their outflow --- 
caused by vehicles that leave the downstream front --- can be either free flow 
or synchronised flow. Typical values for this outflow range from 1500 to 2000 
vehicles/hour/lane \cite{KERNER:04}. The average flow rate within such a 
wide-moving jam can be almost zero, meaning that vehicles continuously encounter 
stop-and-go waves.

Related to the wild scatter in the ($k$,$q$) diagram of three-phase traffic 
theory, is the microscopic behaviour of individual vehicles. The explanation 
given by Kerner and Klenov, is that vehicles in synchronised flow do not assume 
a fixed preferred distance to their direct frontal leader, but rather accept a 
certain \emph{range of distances}. Within this range, drivers have both the 
tendency to over-accelerate when they think there is the ability to overtake, 
and the tendency for drivers to adjust their speed to that of their leader, when 
this overtaking can not be fulfilled \cite{KERNER:03,KERNER:04}.

			\subsubsection{Transitions towards a wide-moving jam}
			\label{sec:TFT:FSJTransitions}

The breakdown of traffic from the free-flow to the wide-moving jam state, is 
nearly always characterised by two successive $F \rightarrow S$ and $S 
\rightarrow J$ transitions, between free flow and synchronised flow, and 
synchronised flow and wide-moving jam respectively. In the first stage, a state 
of free flow changes to synchronised flow by the $F \rightarrow S$ transition. 
Central to the idea of this phase transition, is the fact that there is no 
explicit need for an external disturbance for its occurrence. A sufficiently 
large (i.e., \emph{supercritical}) internal disturbance inside the traffic 
stream (e.g., a lane change) causes a \emph{nucleation effect} that instigates 
the $F \rightarrow S$ transition. Once it has set in, the onset of congestion is 
accompanied by a sharp drop in the mean vehicle speed. During the second stage, 
a set of narrow-moving jams can grow inside the tempo-spatial region of 
synchronised flow. A narrow-moving jam is different from a wide-moving jam, in 
that vehicles typically do not on average come to a full stop inside the jam. 
But, due to a compression of synchronised flow (an effect termed the \emph{pinch 
effect}), these narrow-moving jams can coalesce into a wide-moving jam, thereby 
completing the cascade of the $F \rightarrow S \rightarrow J$ transition, 
resulting in stop-and-go traffic \cite{KERNER:98}.

With respect to the flow versus density diagram in 
\figref{fig:TFT:KernerDensityFlow}, it can be seen that the line $J$ actually 
divides the region of synchronised flow in two parts. Points that lie underneath 
this line, characterise stable traffic states where no $S \rightarrow J$ 
transition can occur. Points above the line $J$ however, characterise metastable 
traffic states, meaning that sufficiently large disturbances can trigger a $S 
\rightarrow J$ transition \cite{KNOSPE:02c,KERNER:04}.

Note that the direct $F \rightarrow J$ transition between free flow and 
wide-moving jam can also occur, but it has a very small probability, i.e., the 
critical perturbation needed, is much higher than that of the frequently 
occurring $F \rightarrow S$ transition between free flow and synchronised flow. 
So in general, wide-moving jams do not emerge spontaneously in free flow, but a 
situation where such a transition may occur, is when an off-ramp gets filled 
with slow-moving vehicles. This results in a local obstruction at the motorway's 
lane directly adjacent to the off-ramp, which can cause a local breakdown of the 
upstream traffic, resulting in a wide-moving jam. Finally, it is important to 
distinguish the nature of this transition from that of the $F \rightarrow S$ 
transition: the former is a transition \emph{induced} by an \emph{external 
disturbance} of the local traffic flow, whereas the latter is considered as a 
\emph{spontaneous} transition due to an \emph{internal disturbance} within the 
local traffic flow (e.g., a lane change) \cite{KERNER:04}.

			\subsubsection{From descriptions to simulations}
			\label{sec:TFM:3PTFromDescriptionsToSimulations}

As Kerner himself describes his three-phase theory, it is a \emph{qualitative} 
theory. In essence, it gives no explanation of \emph{why} certain transitions 
occur, as it only \emph{describes} them \cite{KERNER:04}. However, several 
exemplary microscopic traffic flow models have already been developed (i.e., 
treating all vehicles and their interactions individually). These models can 
reproduce the different empirical tempo-spatial patterns described by Kerner's 
theory. As examples, we mention two models based on cellular automata: a first 
attempt was made by Knospe et al., who developed a model that takes into account 
a driver's reaction to the brake-lights of his direct frontal leader 
\cite{KNOSPE:00b}. Kerner et al. refined this approach by extending it; their 
work resulted in a family of models based on the notion of a 
\emph{synchronisation distance} for individual vehicles; they are commonly 
called the KKW-models (from its three authors, Kerner, Klenov, and Wolf) 
\cite{KERNER:02}.

The theory can describe most of the encountered tempo-spatial features of 
congested traffic. And at the moment, successful microscopic models have been 
developed, but the work is not yet over: an important challenge that remains for 
theoreticians, is the mathematical derivation of a consistent macroscopic theory 
(i.e., one that treats traffic at a more aggregate level as a continuum) 
\cite{KERNER:04}. In pursuit of such a model, Kim incorporated Kerner's traffic 
regimes into a broader framework, encompassing six different possible states: 
the transitions between these states are tracked with a modified macroscopic 
model that uses concepts from fuzzy logic theory \cite{KIM:02}.

		\subsection{Theories of traffic breakdown}
		\label{sec:TFT:TheoriesOfTrafficBreakdown}

A central question that is often asked in the field of traffic flow theory, is 
the following: \emph{``What causes congestion~?''} Clearly, the answer to this 
question should be a bit more detailed than the obvious \emph{``Because there 
too many vehicles on the road~!''} With respect to the phase transitions that 
signal a breakdown of the traffic flow, various --- seemingly contradicting --- 
theories exist. Are they merely a matter of belief, or can they be rigourously 
`proven'~? Opinions are divided, but nowadays, two qualitatively different 
mainstream theories exist, attributed to different schools of thought 
\cite{BUDIANSKY:00,MAERIVOET:04c,TAMPERE:04}:

\textbf{The European (German) school}\\
In the early seventies, Treiterer and Meyers performed some aerial observations 
of a platoon of vehicles. As they constructed individual vehicle trajectories, 
they could observe a growing instability in the stream of vehicles, leading to 
an apparently emerging \emph{phantom jam} (i.e., a jam `out of nothing') 
\cite{TREITERER:74}.

Some twenty years later, in the mid-nineties, Kerner and Konh\"auser made 
detailed studies of traffic flow measurements, obtained at various detector 
stations along German motorways. Their findings indicated that phantom jams 
seemed to emerge in regions of unstable traffic flow \cite{KERNER:94}. This 
stimulated Kerner and Rehborn to further research efforts directed towards the 
behaviour of propagating jams \cite{KERNER:96,KERNER:96b}. They proposed a 
different set of traffic flow regimes, culminating in what is now called 
\emph{three-phase traffic theory} (see section \ref{sec:TFT:Kerners3PT} for more 
details) \cite{KERNER:97,KERNER:98,KERNER:04}. The main idea supported by 
followers of this school of researchers, is that traffic jams \emph{can} 
spontaneously emerge, without necessarily having an infrastructural reason 
(e.g., on-ramps, incidents, \ldots)\footnote{But note that bottleneck-induced 
traffic flow breakdowns are not excluded by the theory of Kerner et al.}. In 
dense enough traffic, phase transitions from the free-flow to the 
synchronised-flow regime can occur, after which a local instability such as 
e.g., a lane change can grow (the so-called \emph{pinch effect}), triggering a 
stable jam leading to stop-and-go behaviour \cite{KERNER:98}. Kerner's 
three-phase theory stands out as an archetypical example of these modern views. 
But although his theory has, in our opinion, been worked out well enough, he 
more than frequently encounters harsh criticisms when conveying it to most 
audiences (perhaps the main cause for this human behaviour is the fact that 
Kerner always mentions the same view, i.e., \emph{``all existing traffic flow 
theories are wrong''}).

Inspired by Kerner's work, Helbing et al. gave in 1999 an extended treatise on 
the different types of congestion patterns that can be observed in the vicinity 
of spatial inhomogeneities (e.g., on-ramps). Their work resulted in a universal 
phase diagram, containing a whole plethora of patterns of congested traffic 
states (called \emph{homogeneously congested traffic} -- HCT, \emph{oscillatory 
congested traffic} -- OCT, \emph{triggered stop-and-go traffic} -- TSG, 
\emph{pinned localised cluster} -- PLC, and \emph{moving localised cluster} -- 
MLC), each one having unique characteristics \cite{HELBING:99d}\footnote{In 
addition, they also provided a link with Kerner's three-phase theory, whereby 
synchronised flow can correspond to HCT, OCT, or PLC, and moving jams can 
correspond to TSG or MLC states \cite{HELBING:02}.}. In that same year, Lee et 
al. studied the patterns that emerge at on-ramps, thereby agreeing with the 
findings of Helbing et al. \cite{LEE:99}. As the previous research into 
congestion patterns was largely based on the use of analytical traffic flow 
models and computer simulations, the need for validation with empirical data 
grew. In 2000, the work of Treiber et al. among others, proved the existence of 
the previously mentioned congestion patterns \cite{TREIBER:00}.

At this point, it is noteworthy to mention the seminal work of Nagel and 
Schreckenberg \cite{NAGEL:92}, who in 1992 developed a model that describes 
traffic flows in which local jams can form spontaneously. As many variations on 
this model have been proposed, later work also focussed on the stability of 
traffic flows in these models, e.g., the work of Jost and Nagel 
\cite{JOST:03}.\\

\textbf{The Berkeley school}\\
Including names such as the late Newell, Daganzo, Bertini, Cassidy, Mu\~noz, 
\ldots, the `Berkeley school' (University of California) supports the theory 
that all congestion is strictly induced by bottlenecks. The hypothesis holds for 
both recurrent and, in the case of an incident, non-recurrent congestion.

The main starting point states that there is \emph{always} a `geometrical' 
explanation for the breakdown. This explanation is based on the presence of road 
inhomogeneities such as on- and off-ramps, tunnels, weaving areas, lane drops, 
sharp bends, elevations, \ldots Once a jam occurs due to such a (temporary) 
bottleneck, it does not dissipate immediately; as a result, drivers can wonder 
why they enter and exit a congestion wave, without there being an apparent 
reason for its presence (since it happened earlier and the cause e.g., an 
incident, already got cleared). Daganzo uses this line of reasoning as an 
explanation for the dismissal of phantom jams \cite{DAGANZO:02c}.

The school uses a specific terminology with respect to bottlenecks (being road 
inhomogeneities). Two qualitatively different regimes exist: the \emph{free-flow 
regime} and the \emph{queued regime}. The latter occurs when a bottleneck 
becomes \emph{active}, which will result in a queue growing upstream of the 
bottleneck while a free-flow regime exists downstream. The \emph{bottleneck 
capacity} is then defined as the maximum \emph{sustainable} flow downstream 
(which is different from the maximum flow that can be observed prior to the 
bottleneck's activation).

The location of these bottlenecks has some peculiarities involved: one of them 
is the concept of a \emph{capacity funnel} \cite{BUCKLEY:74}. It assumes that 
drivers are at times more alert, e.g., when they are driving on a motorway and 
nearing an on-ramp in rather dense traffic conditions \cite{ZHANG:01b}. This 
impels them to accept shorter headways, so they are driving closely behind each 
other at a relatively high speed. Once they have passed the on-ramp's location, 
they tend to relax, resulting in larger headways. The effect is that the 
bottleneck's \emph{actual} position is located more downstream.

Shortly after the publication of Kerner and Rehborn's findings about the 
peculiar phase transitions that seemed to occur on German motorways, Daganzo et 
al. provided a swift response where they stated that the occurring phase 
transitions \emph{could} also be caused by bottlenecks in a predictable way 
\cite{DAGANZO:99b}. They implied that no spontaneously emerging traffic jams are 
suggested, and that the observed traffic data from both German and North 
American motorways did not contradict their own statements about the cause of 
the phase transitions \cite{NAGEL:05}. In short, the subtle difference between 
their work and that of Kerner and Rehborn, is that instabilities in the traffic 
stream are the \emph{result} and not the cause of the queues that emerge at 
active bottlenecks. With respect to a spontaneous breakdown of traffic flow at 
on- and off-ramps (i.e., bottlenecks), Daganzo also states that this can be 
explained using a simple traffic flow model operating under the assumption of a 
too high inflow from the on-ramp or a caused by blocking of the off-ramp 
\cite{DAGANZO:96}.

The studies undertaken by this school, are heavily based on the researchers' use 
of cumulative plots and elegantly simple traffic flow models, as opposed to the 
classic methodology that investigates time series of recorded counts and speeds. 
As stated earlier (see section \ref{sec:TFT:ObliqueCumulativePlots}), some 
recent examples include the work of Mu\~noz and Daganzo 
\cite{MUNOZ:00b,MUNOZ:00,MUNOZ:02b,MUNOZ:03b} and Cassidy and Bertini 
\cite{CASSIDY:99,BERTINI:03}.

Recently, Tamp\`ere argued that both theories, as enunciated by the two schools, 
are not entirely contradictory. His statement is based on the fact that the 
mechanisms behind the bottleneck-induced breakdown and spontaneous breakdown are 
approximately the same, only differing in the \emph{probability} of such a 
breakdown (which is related to the instability of a traffic flow) 
\cite{TAMPERE:04}.\\

\sidebar{
	In our view, both theories are sufficiently different, \emph{but compatible}, 
	in that the first school elaborately describes traffic flow breakdown more or 
	less as having an inherently \emph{probabilistic nature}, whereas the second 
	school treats breakdown a strict \emph{deterministic process}. The former 
	introduces a complex variety of congestion patterns, while the latter 
	primarily focusses on an elegantly simple description of traffic flow 
	breakdown. Even more characteristically, is the observation that most adepts 
	of the European school, \emph{inherently need stochasticity in the models} in 
	order to produce their sought phantom traffic jams (note that notwithstanding 
	the fact that stochastic models are in a strict sense also deterministic, we 
	nevertheless adopt in this dissertation, the convention that deterministic 
	means `non-stochastic'). Our argument is in a way also supported by Nagel and 
	Nelson, who state that the purpose of the traffic flow model (e.g., the effect 
	of moving bottlenecks versus predicting mean traffic behaviour) decide whether 
	or not stochasticity in the model is required \cite{NAGEL:05}. 
}\\

\sidebar{
	Furthermore, there might be some room for stochasticity in the Berkeley models 
	after all, with the work of Laval which suggests that (disruptive) lane 
	changes form the main cause for instabilities in a traffic stream 
	\cite{LAVAL:04}. Deciding which school is right, is therefore in our opinion a 
	matter of personal taste, but in the end, we agree with Daganzo when he states 
	that research into bottleneck behaviour is the most important in the context 
	of traffic flow theory \cite{DAGANZO:99}.
}\\

	\section{Conclusions}

In this paper, an extensive account was given, detailing several aspects related 
to the description of traffic flows. Most importantly, we have introduced a 
nomenclature convention, built upon a consistent set of notations. Our 
discussion of traffic flow characteristics centred around the space and time 
headways as microscopic characteristics, with densities and flows as their 
macroscopic counterparts. Several noteworthy highlights are the technique of 
oblique cumulative plots and the derivation of travel times based on these 
plots. A finally large part of this paper reviewed some of the relations between 
traffic flow characteristics, i.e., the fundamental diagrams, and clarified some 
of the different points of view adopted by the traffic engineering community.

\appendix
%

\section{Glossary of terms}
\label{appendix:Glossary}

	\subsection{Acronyms and abbreviations}

\begin{tabular}{ll}
	4SM            & four step model\\
	AADT           & annual average daily traffic\\
	ABM            & activity-based modelling\\
	ACC            & adaptive cruise control\\
	ACF            & average cost function\\
	ADAS           & advanced driver assistance systems\\
	AIMSUN2        & Advanced Interactive Microscopic\\
	               & Simulator for Urban and Non-Urban\\
	               & Networks\\
	AMICI          & Advanced Multi-agent Information and\\
	               & Control for Integrated multi-class traffic\\
	               & networks\\
	AON            & all-or-nothing\\
	ASDA           & Automatische StauDynamikAnalyse\\
	ASEP           & asymmetric simple exclusion process\\
	ATIS           & advanced traveller information systems\\
	ATMS           & advanced traffic management systems\\
	BCA            & Burgers cellular automaton\\
	BJH            & Benjamin, Johnso, and Hui\\
	BJH-TCA        & Benjamin-Johnson-Hui traffic cellular\\
	               & automaton\\
	BL-TCA         & brake-light traffic cellular automaton\\
	BML            & Biham, Middleton, and Levine\\
	BML-TCA        & Biham-Middleton-Levine traffic cellular\\
	               & automaton\\
	BMW            & Beckmann, McGuire, and Winsten\\
	BPR            & Bureau of Public Roads\\
\end{tabular}

\begin{tabular}{ll}
	CA             & cellular automaton\\
	CA-184         & Wolfram's cellular automaton rule 184\\
	CAD            & computer aided design\\
	CBD            & central business district\\
	CFD            & computational fluid dynamics\\
	CFL            & Courant-Friedrichs-Lewy\\
	ChSch-TCA      & Chowdhury-Schadschneider traffic\\
	               & cellular automaton\\
	CLO            & camera Linkeroever\\
	CML            & coupled map lattice\\
	CONTRAM        & CONtinuous TRaffic Assignment\\
	               & Model\\
	COMF           & car-oriented mean-field theory\\
	CPM            & computational process models\\
	CTM            & cell transmission model\\
	DDE            & delayed differential equation\\
	DFI-TCA        & deterministic Fukui-Ishibashi traffic\\
	               & cellular automaton\\
	DGP            & dissolving general pattern\\
	DLC            & discretionary lane change\\	
	DLD            & double inductive loop detector\\
	DNL            & dynamic network loading\\
	DRIP           & dynamic route information panel\\
	DTA            & dynamic traffic assignment\\
	DTC            & dynamic traffic control\\
	DTM            & dynamic traffic management\\
	DUE            & deterministic user equilibrium\\
	DynaMIT        & Dynamic network assignment for the\\
	               & Management of Information to\\
	               & Travellers\\
	DYNASMART      & DYnamic Network Assignment-\\
	               & Simulation Model for Advanced\\
	               & Roadway Telematics\\
	ECA            & elementary cellular automaton\\
	EP             & expanded congested pattern\\
	ER-TCA         & Emmerich-Rank traffic cellular\\
	               & automaton\\
	FCD            & floating car data\\
	FDE            & finite difference equation\\
	FIFO           & first-in, first-out\\
	FOTO           & Forecasting of Traffic Objects\\
	GETRAM         & Generic Environment for TRaffic\\
	               & Analysis and Modeling\\
	GHR            & Gazis-Herman-Rothery\\
	GIS            & geographical information systems\\
	GNSS           & Global Navigation Satellite System\\
	               & (e.g., Europe's Galileo)\\
	GoE            & Garden of Eden state\\
	GP             & general pattern\\
	GPRS           & General Packet Radio Service\\
	GPS            & Global Positioning System\\
	               & (e.g., USA's NAVSTAR)\\
	GRP            & generalised Riemann problem\\
	GSM            & Groupe Sp\'eciale Mobile\\
	GSMC           & Global System for Mobile\\
	               & Communications\\
	HAPP           & household activity pattern problem\\
	HCM            & Highway Capacity Manual\\
	HCT            & homogeneously congested traffic\\
	HDM            & human driver model\\
	HKM            & human-kinetic model\\
	HRB            & Highway Research Board\\
\end{tabular}

\begin{tabular}{ll}
	HS-TCA         & Helbing-Schreckenberg traffic cellular\\
	               & automaton\\
	ICC            & intelligent cruise control\\
	IDM            & intelligent driver model\\
	INDY           & INteractive DYnamic traffic assignment\\
	ITS            & intelligent transportation systems\\
	IVP            & initial value problem\\
	JDK            & Java\trademark Development Kit\\
	KKT            & Karush-Kuhn-Tucker\\
	KKW-TCA        & Kerner-Klenov-Wolf traffic cellular\\
	               & automaton\\
	KWM            & kinematic wave model\\
	LGA            & lattice gas automaton\\
	LOD            & level of detail\\
	LOS            & level of service\\
	LSP            & localised synchronised-flow pattern\\
	LTM            & link transmission model\\
	LWR            & Lighthill, Whitham, and Richards\\
	MADT           & monthly average daily traffic\\
	MC-STCA        & multi-cell stochastic traffic cellular\\
	               & automaton\\
	MesoTS         & Mesoscopic Traffic Simulator\\
	MFT            & mean-field theory\\
	MITRASIM       & MIcroscopic TRAffic flow SIMulator\\
	MITSIM         & MIcroscopic Traffic flow SIMulator\\
	MIXIC          & Microscopic model for Simulation of\\
	               & Intelligent Cruise Control\\
	MLC            & mandatory lane change\\	
	               & moving localised cluster\\	
	MOE            & measure of effectiveness\\
	MPA            & matrix-product ansatz\\
	MPCF           & marginal private cost function\\
	MSA            & method of successive averages\\
	MSCF           & marginal social cost function\\
	MSP            & moving synchronised-flow pattern\\
	MT             & movement time\\
	MUC-PSD        & multi-class phase-space density\\
	NaSch          & Nagel and Schreckenberg\\
	NAVSTAR        & Navigation Satellite Timing and Ranging\\
	NCCA           & number conserving cellular automaton\\
	NSE            & Navier-Stokes equations\\
	OCT            & oscillatory congested traffic\\
	OD             & origin-destination\\
	ODE            & ordinary differential equation\\
	OSS            & Open Source Software\\
	OVF            & optimal velocity function\\
	OVM            & optimal velocity model\\
	Paramics       & Parallel microscopic traffic simulator\\
	PATH           & California Partners for Advanced Transit\\
	               & and Highways\\
	               & Program on Advanced Technology for\\
	               & the Highway\\
	PCE            & passenger car equivalent\\
	PCU            & passenger car unit\\
	PDE            & partial differential equation\\
	PELOPS         & Program for the dEvelopment of\\
	               & Longitudinal micrOscopic traffic\\
	               & Processes in a Systemrelevant\\
	               & environment\\
	PeMS           & California Freeway Performance\\
	               & Measurement System\\
	PHF            & peak hour factor\\
\end{tabular}

\begin{tabular}{ll}
	PLC            & pinned localised cluster\\
	pMFT           & paradisiacal mean-field theory\\
	PRT            & perception-reaction time\\
	PSD            & phase-space density\\
	PW             & Payne-Whitham\\
	QoS            & quality of service\\
	SFI-TCA        & stochastic Fukui-Ishibashi traffic\\
	               & cellular automaton\\
	Simone         & Simulation model of Motorways with\\
	               & Next generation vehicles\\
	SLD            & single inductive loop detector\\
	SMARTEST       & Simulation Modelling Applied to Road\\
	               & Transport European Scheme Tests\\
	SMS            & space-mean speed\\
	SOC            & self-organised criticality\\
	SOMF           & site-oriented mean-field theory\\
	SP             & synchronised-flow pattern\\
	SSEP           & symmetric simple exclusion process\\
	STA            & static traffic assignment\\
	STCA           & stochastic traffic cellular automaton\\
	STCA-CC        & stochastic traffic cellular automaton\\
	               & with cruise control\\
	SUE            & stochastic user equilibrium\\
	SUMO           & Simulation of Urban MObility\\
	T$^{\two}$-TCA & Takayasu-Takayasu traffic cellular\\
	               & automaton\\
	TASEP          & totally asymmetric simple exclusion\\
	               & process\\
	TCA            & traffic cellular automaton\\
	TDF            & travel demand function\\
	TMC            & Traffic Message Channel\\
	TMS            & time-mean speed\\
	TOCA           & time-oriented traffic cellular\\
	               & automaton\\
	TRANSIMS       & TRansportation ANalysis and SIMulation\\
	               & System\\
	TRB            & Transportation Research Board\\
	TSG            & triggered stop-and-go traffic\\
	UDM            & ultra-discretisation method\\
	UMTS           & Universal Mobile Telecommunications\\
	               & System\\
	VDR-TCA        & velocity-dependent randomisation traffic\\
	               & cellular automaton\\
	VDT            & total vehicle distance travelled\\
	VHT            & total vehicle hours travelled\\
	VMS            & variable message sign\\
	VMT            & total vehicle miles travelled\\
	VOT            & value of time\\
	WSP            & widening synchronised-flow pattern\\
	WYA            & whole year analysis\\
\end{tabular}

	\subsection{List of symbols}

\renewcommand{\arraystretch}{1.2}

\begin{tabular}{ll}
	$a_{i}$                                & the acceleration of vehicle $i$\\
	$C$                                    & the number of substreams in a traffic flow\\
	$dx$                                   & a single infinitesimal location in space\\
	$dt$                                   & a single infinitesimal instant in time\\
	$\eta$                                 & the efficiency of a road section\\
	                                       & (according to Chen et al, \cite{CHEN:01})\\
\end{tabular}

\begin{tabular}{ll}
	$E$                                    & the efficiency of a road section\\
	                                       & (according to Brilon, \cite{BRILON:00})\\
	$F$                                    & the free-flow curve in three-phase traffic theory\\
	$g_{s_{i}}$                            & the space gap of vehicle $i$\\
	$g_{s_{i}}^{l,b}$                      & the space gap at the left-back of vehicle $i$\\
	$g_{s_{i}}^{l,f}$                      & the space gap at the left-front of vehicle $i$\\
	$g_{s_{i}}^{r,b}$                      & the space gap at the right-back of vehicle $i$\\
	$g_{s_{i}}^{r,f}$                      & the space gap at the right-front of vehicle $i$\\
	$g_{t_{i}}$                            & the time gap of vehicle $i$\\
	$g_{t_{i}}^{l,b}$                      & the time gap at the left-back of vehicle $i$\\
	$g_{t_{i}}^{l,f}$                      & the time gap at the left-front of vehicle $i$\\
	$g_{t_{i}}^{r,b}$                      & the time gap at the right-back of vehicle $i$\\
	$g_{t_{i}}^{r,f}$                      & the time gap at the right-front of vehicle $i$\\
	$\overline h_{s}$                      & the average space headway\\
	$h_{s_{i}}$                            & the space headway of vehicle $i$\\
	$h_{s_{i}}^{l,b}$                      & the space headway at the left-back of vehicle $i$\\
	$h_{s_{i}}^{l,f}$                      & the space headway at the left-front of vehicle $i$\\
	$h_{s_{i}}^{r,b}$                      & the space headway at the right-back of vehicle $i$\\
	$h_{s_{i}}^{r,f}$                      & the space headway at the right-front of vehicle $i$\\
	$\overline h_{t}$                      & the average time headway\\
	$h_{t_{i}}$                            & the time headway of vehicle $i$\\
	$h_{t_{i}}^{l,b}$                      & the time headway at the left-back of vehicle $i$\\
	$h_{t_{i}}^{l,f}$                      & the time headway at the left-front of vehicle $i$\\
	$h_{t_{i}}^{r,b}$                      & the time headway at the right-back of vehicle $i$\\
	$h_{t_{i}}^{r,f}$                      & the time headway at the right-front of vehicle $i$\\
	$J$                                    & the wide moving jam line $J$ in three-phase\\
	                                       & traffic theory\\
	$k$                                    & the density\\
	$k_{c}$                                & the density of the $c$-th substream in a\\
	                                       & traffic flow\\
	$k_{c}$                                & the critical density\\
	$k_{\text{crit}}$                      & the critical density\\
	$k_{j}$                                & the jam density\\
	$k_{\text{jam}}$                       & the jam density\\
	$k_{\text{max}}$                       & the jam density\\
	$k_{l}$                                & the density in lane $l$\\
	$k_{\text{out}}$                       & the density associated with the queue discharge\\
	                                       & capacity\\
	$k(t)$                                 & the density at time $t$\\
	$K$                                    & the length of a measurement region\\
	                                       & (i.e., a certain road section)\\
	$K_{\text{ld}}$                        & the length of a detection zone\\
	$\overline l$                          & the average length of a vehicle\\
	$l_{i}$                                & the length of vehicle $i$\\
	$L$                                    & the number of lanes on a road\\
	$N$                                    & the number of vehicles in a measurement region\\
	$N_{l}$                                & the number of vehicles in the measurement\\
	                                       & region in lane $l$\\
	$N_{l}(t)$                             & the number of vehicles in the measurement\\
	                                       & region in lane $l$ at time $t$\\
	$N(t)$                                 & a cumulative count function\\
\end{tabular}

\begin{tabular}{ll}
	$\widetilde N(t)$                      & a smooth approximation of $N(t)$\\
	$\overline o_{t}$                      & the average on-time of a set of vehicles\\
	$o_{t_{i}}$                            & the on-time of vehicle $i$\\
	$o_{t_{i,l}}$                          & the on-time of vehicle $i$ in lane $l$\\
	$q$                                    & the flow\\
	$\overline q_{|\text{15}}$             & the peak flow rate during one quarter hour\\
	                                       & within an hour\\
	$\overline q_{|\text{60}}$             & the average flow during the hour with the\\
	                                       & maximum flow in one day\\
	$q_{b}$                                & a background flow\\
	$q_{c}$                                & the flow of the $c$-th substream in a traffic flow\\
	$q_{c}$                                & the capacity flow\\
	$q_{\text{cap}}$                       & the capacity flow\\
	$q_{e}(k)$                             & an equilibrium relation between the flow and\\
	                                       & the density\\
	$q_{l}$                                & the flow in lane $l$\\
	$q_{\text{max}}$                       & the capacity flow\\
	$q_{\text{out}}$                       & the outflow from a (wide moving) jam,\\
	                                       & the queue discharge capacity\\
	$q(t)$                                 & the flow at time $t$\\
	$\rho$                                 & the occupancy\\
	$\rho_{i}$                             & the occupancy time of vehicle $i$\\
	$\rho_{l}$                             & the occupancy in lane $l$\\
	$R_{s}$                                & a spatial measurement region at a fixed time\\
	                                       & instant\\
	$R_{t}$                                & a temporal measurement region at a fixed\\
	                                       & location\\
	$R_{t,s}$                              & a general measurement region\\
	$\sigma_{s}^{\two}$                    & the statistical sample variance of the space-\\
	                                       & mean speed\\
	$\sigma_{t}^{\two}$                    & the statistical sample variance of the time-\\
	                                       & mean speed\\
	$S$                                    & the synchronised-flow region in three-phase\\
	                                       & traffic theory\\
	$\tau_{i}$                             & the reaction time of vehicle $i$'s driver\\
	$t$                                    & a time instant\\
	$T_{i}$                                & the travel time of vehicle $i$\\
	$T_{\text{mp}}$                        & the duration of a measurement period\\
	$T(t_{\zero})$                         & the experienced dynamic travel time, starting at\\
	                                       & time instant $t_{\zero}$\\
	$\widetilde{T}(t_{\zero})$             & the experienced instantaneous travel time,\\
	                                       & starting at time instant $t_{\zero}$\\
	$\overline v_{c}$                      & the capacity-flow speed\\
	$\overline v_{\text{cap}}$             & the capacity-flow speed\\
	$\overline v_{\text{ff}}$              & the free-flow speed\\
	$v_{i}$                                & the speed of vehicle $i$\\
	$v_{i,l}$                              & the speed of vehicle $i$ in lane $l$\\
	$v_{i,l}(t)$                           & the speed of vehicle $i$ in lane $l$ at time $t$\\
	$v_{\text{max}}$                       & the maximum allowed speed (e.g., by an\\
	                                       & imposed speed limit)\\
	$\overline v_{s}$                      & the space-mean speed\\
\end{tabular}

\begin{tabular}{ll}
	$\overline v_{s_{c}}$                  & the space-mean speed of the $c$-th substream\\
	$\overline v_{s_{e}}(\overline h_{s})$ & an equilibrium relation between the\\
	                                       & SMS and the average space headway\\
	$\overline v_{s_{e}}(k)$               & an equilibrium relation between the\\
	                                       & SMS and the density\\
	$\overline v_{s_{e}}(q)$               & an equilibrium relation between the\\
	                                       & SMS and the flow\\
	$\overline v_{\text{sust}}$            & the sustained speed during a period of high\\
	                                       & flow\\
	$\overline v_{t}$                      & the space-mean speed\\
	$\overline v_{t_{c}}$                  & the time-mean speed of the $c$-th substream\\
	$v(t,x)$                               & the local instantaneous vehicle speed at time\\
	                                       & instant $t$ and location $x$\\
	$w$                                    & the characteristic/kinematic wave speed (of a\\
	                                       & wide moving jam)\\
	$x_{i}$                                & the longitudinal position of vehicle $i$\\
	$X_{i}$                                & the distance travelled by vehicle $i$\\
\end{tabular}

\section*{Acknowledgements}

Dr. Bart De Moor is a full professor at the Katholieke Universiteit Leuven, Belgium.
\noindent
Our research is supported by:
\textbf{Research Council KUL}: GOA AMBioRICS, several PhD/post\-doc
\& fellow grants,
\textbf{Flemish Government}:
\textbf{FWO}: PhD/post\-doc grants, projects, G.0407.02 (support vector machines),
G.\-0197.02 (power islands), G.0141.03 (identification and cryptography), G.0491.03
(control for intensive care glycemia), G.0120.\-03 (QIT), G.0452.04 (new quantum algorithms),
G.0499.04 (statistics), G.0211.05 (Nonlinear), research communities (ICCoS, ANMMM, MLDM),
\textbf{IWT}: PhD Grants, GBOU (McKnow),
\textbf{Belgian Federal Science Policy Office}: IUAP P5/\-22 (`Dynamical Systems and
Control: Computation, Identification and Modelling', 2002-2006), PODO-II (CP/40:
TMS and Sustainability),
\textbf{EU}: FP5-Quprodis, ERNSI,
\textbf{Contract Research/agreements}: ISMC/IPCOS, Data4s,TML, Elia, LMS,
Mastercard.

\bibliography{paper}

\end{document}